\pdfoutput=1
\documentclass[AMA,LATO1COL]{WileyNJDv5} 

\articletype{Review Article}%

\received{Date Month Year}
\revised{Date Month Year}
\accepted{Date Month Year}
\journal{Journal}
\volume{00}
\copyyear{2025}
\startpage{1}

\usepackage[T1]{fontenc}
\usepackage{helvet}  

\usepackage[utf8]{inputenc}
\usepackage{lmodern}
\renewcommand{\textendash}{--} 

\usepackage{amssymb}
\usepackage{makecell}
\usepackage{multirow}
\usepackage{graphicx}
\usepackage{booktabs}
\usepackage{array}
\usepackage{xcolor}
\usepackage{color, colortbl}
\usepackage{longtable} 
\usepackage{pdflscape}
\usepackage{hhline}
\usepackage{comment}
\usepackage{caption}
\usepackage{subcaption}
\usepackage{lipsum}
\usepackage{orcidlink}
\usepackage[many]{tcolorbox}    	
\usepackage{listings}
\usepackage{enumitem}
\usepackage{csquotes}

\usepackage[first=0,last=9]{lcg}
\usepackage[inkscapeformat=png]{svg}

\setlist[description]{style=nextline}

\definecolor{antiquewhite}{rgb}{0.98, 0.92, 0.84}
\definecolor{anti-flashwhite}{rgb}{0.92, 0.92, 0.92}
\definecolor{Gray}{gray}{0.85}
\definecolor{darkgray}{rgb}{0.75, 0.75, 0.75}
\definecolor{dimgray}{rgb}{0.58, 0.58, 0.58}
\definecolor{battleshipgrey}{rgb}{0.60, 0.60, 0.61}
\definecolor{davysgrey}{rgb}{0.43, 0.43, 0.43}

\newcolumntype{a}{>{\columncolor{anti-flashwhite}}c}
\newcolumntype{d}{>{\columncolor{Gray}}c}
\newcolumntype{e}{>{\columncolor{darkgray}}c}
\newcolumntype{f}{>{\columncolor{dimgray}}c}
\newcolumntype{i}{>{\columncolor{battleshipgrey}}c}
\newcolumntype{h}{>{\columncolor{davysgrey}}c}

\usepackage{hyperref}

\hypersetup{
    colorlinks=true,
    linkcolor=blue,
    filecolor=blue,      
    urlcolor=blue,
    pdfauthor={Edi Sutoyo, Andrea Capiluppi}
    }
\newtcolorbox[auto counter]{somebox}[1][]{arc=5pt,auto outer arc,left=1pt,boxsep=0.5pt,boxrule=0.5pt,width=\columnwidth,right=1pt, #1}

\begin{document}

\title{Self-Admitted Technical Debt Detection Approaches: A Decade Systematic Review}



\author[1,2]{Edi Sutoyo} 

\author[1]{Andrea Capiluppi} 


\authormark{SUTOYO and CAPILUPPI} 
\titlemark{SELF-ADMITTED TECHNICAL DEBT DETECTION APPROACHES: A DECADE SYSTEMATIC REVIEW}



\address[1]{\orgdiv{Bernoulli Institute for Mathematics, Computer Science and Artificial Intelligence}, \orgname{University of Groningen}, \orgaddress{\state{Groningen}, \country{The Netherlands}}}

\address[2]{\orgdiv{Department of Information Systems}, \orgname{Telkom University}, \orgaddress{\state{Bandung}, \country{Indonesia}}}


\corres{Edi Sutoyo, Bernoulli Institute, University of Groningen, Nijenborgh 9, 9747 AG. \email{e.sutoyo@rug.nl}}


\abstract[Abstract]{
Technical debt (TD) refers to the long-term costs associated with suboptimal design or code decisions in software development, often made to meet short-term delivery goals. Self-Admitted Technical Debt (SATD) occurs when developers explicitly acknowledge these trade-offs in the codebase, typically through comments or annotations. SATD detection has become an increasingly important research area, particularly with the rise of learning-based techniques that aim to streamline SATD detection.

This systematic literature review provides a comprehensive analysis of SATD detection approaches published between 2014 and early 2025, focusing on the evolution of techniques from heuristic-based techniques to more advanced ML, DL, and Transformer-based models. It examines key trends in SATD detection methodologies and tools, evaluates the effectiveness of different approaches using metrics like precision, recall, and F1 score, and highlights the primary challenges in this domain, including dataset heterogeneity, model generalizability, and explainability.

The findings reveal that while early heuristic-based techniques laid the foundation for SATD detection, more recent advancements in DL and Transformer models have significantly improved detection accuracy. However, challenges remain in scaling these models for broader industrial adoption. This review offers insights into current research gaps and provides directions for future work, aiming to improve the robustness and practicality of SATD detection tools.
}


\jnlcitation{\cname{%
\author{Sutoyo E.}, and
\author{Capiluppi A}}.
\ctitle{Self-Admitted Technical Debt Detection Approaches: A Decade Systematic Review} \cjournal{\it J Softw Evol Proc.} \cvol{2024;00(00):x--xx}.}

\maketitle



\section{Introduction}\label{sec:Introduction}
Business needs often require developers to rapidly deliver new products and features to reach the market quickly. It is important for developers to carefully balance design and implementation to seamlessly add features. However, in the business or corporate world, schedules and limited resources often hinder developers from writing perfect code before incorporating it into products \citep{RIOS2018117}. This becomes commonplace when stakeholders cut technical processes to gain momentum, especially in startups \citep{Klinger2011AnDebt, Besker2018EmbracingPerspective}. Tight deadlines often lead to technical debt because developers lack time to consider long-term solutions \citep{Avgeriou2016ManagingEngineering}. 

Technical debt (TD) is often used to describe the trade-off that occurs when an organization prioritizes delivering new features or functionality over maintaining the software in the long term \citep{Cunningham1992TheSystem}.  TD can occur in any software development project and may be caused by a variety of factors, including insufficient resources, time, knowledge, or experience on the part of the development team \citep{Brown2010ManagingSystems, Tom2013AnDebt}.


Among various types of TD, self-admitted technical debt (SATD) has gained considerable attention due to its explicit acknowledgment within the source code comments by developers. SATD refers to instances in which developers, either intentionally or unintentionally, make suboptimal design or implementation choices, despite having a clear understanding of the negative impact these choices will have on the codebase~\citep{Potdar2014AnDebt}. This acknowledgment is not merely an admission of suboptimal practices but a vital piece of information that can guide future development activities. Recognizing SATD is crucial for maintaining software quality and sustainability, as it can accumulate over time and impede future development efforts.

Over the past decade, various approaches to SATD detection have emerged, evolving from heuristic-based techniques to more advanced machine learning (ML), deep learning (DL), and Transformer-based models. These approaches aim to automate SATD detection, facilitate its management, and enable software teams to address it proactively. Alongside these approaches, various methodologies have been proposed to detect SATD, ranging from manual to semi-automated to automated methodologies. However, the diversity of techniques, methodologies, tools, and evaluation metrics used in SATD detection has led to a fragmented landscape, making it challenging for researchers and practitioners to gain a holistic understanding of the field.

To address this fragmentation, we selected a study that aligns closely with our focus on SATD detection. We summarize this work, compare it with our systematic literature review (SLR), and provide a condensed analysis of the key differences. Our SLR distinguishes itself from prior reviews in several key areas:




\begin{description}

\item[Time frame and scope.] Our review spans a decade-long period from January 2014 to April 2025, analyzes 74 studies on SATD detection, and provides the most up-to-date synthesis of SATD research. In contrast, the review by Sierra et al. \citep{Sierra2019ADebt} focuses on a shorter interval (2014\textendash2018) and captures earlier SATD detection efforts. Similarly, Gu et al. \citep{gu2024self} conduct a lightweight review that covers literature until 2022, but primarily focuses on evaluating detection models across different software artifacts rather than providing a broad literature synthesis.

\item[Systematic approach.] We follow the established and rigorous procedures outlined by Kitchenham and Charters \citep{Kitchenham2007GuidelinesEngineering} for conducting systematic reviews. In contrast, Sierra et al. \citep{Sierra2019ADebt} appears to be based on a survey approach, without explicitly detailing how studies are sourced. Gu et al. \citep{gu2024self} conduct an empirical study, but their review methodology does not follow an SLR framework, as it primarily assesses the generalization of SATD models across programming languages and software artifacts.

\item[Comprehensiveness.] In addition to relying on digital library searches, our review employs a bidirectional snowballing process to capture all relevant studies and results, resulting in 74 analyzed papers and making our review broader and more representative of the field. While Sierra et al. \citep{Sierra2019ADebt} also employ snowballing, their review covers a narrower time frame (2014\textendash2018), which may limit the discovery of additional relevant studies. In contrast, Gu et al. \citep{gu2024self} do not explicitly perform snowballing, primarily focusing on evaluating existing SATD detection models rather than conducting a comprehensive literature search.

\item[Depth of analysis.] Prior reviews, such as Sierra et al. \citep{Sierra2019ADebt} and Gu et al. \citep{gu2024self}, do not consistently synthesize reported findings. Our review extends this by examining both methodological characteristics (e.g., degree of automation) and empirical evidence (e.g., reported performance metrics), offering a more integrated perspective on how SATD detection techniques evolve and perform.




\end{description}

In this study, we conduct a systematic literature review on SATD detection, evaluating existing approaches and their effectiveness. The objectives and key contributions of our study are as follows:

\begin{itemize}
\item We consolidate and classify SATD detection approaches, tracing the evolution from heuristic-based methods to ML, DL, and Transformer-based models. This taxonomy provides researchers with a structured view of methodological progress in the field.

\item We compare methodologies in terms of their level of automation (manual, semi-automated, and automated) and synthesize reported performance metrics (precision, recall, and F1 score) across diverse datasets and software systems. This synthesis provides a consolidated overview of the techniques and highlights the best-performing models reported in prior studies.

\item We conduct the first systematic review of SATD detection tools, assessing their availability, capabilities, and reported effectiveness. This analysis reveals persistent gaps between research prototypes and practical tool adoption in industry settings.

\item We identify and synthesize nine recurring challenges in SATD detection (e.g., dataset imbalance, model generalizability and explainability, and adoption barriers), which highlight key opportunities for future research.

\end{itemize}

In addition to the Introduction, the remaining study sections are organized as follows: Section~\ref{section_background} introduces the concept and background, and Section~\ref{section_rm} describes the methodology utilized to conduct the SLR. Sections~\ref{section_results} and~\ref{section_discussion} summarize and discuss the obtained results, while Section~\ref{section_threats} identifies the threats to validity of this study. Finally, Section~\ref{section_conclusion} offers a conclusion.

\section{Background} 
\label{section_background}

This section introduces the key concepts underlying this study, focusing on self-admitted technical debt (SATD) and the role of natural language processing (NLP) techniques that are directly relevant for analyzing the textual artifacts used in SATD detection.



\subsection{Self-Admitted Technical Debt (SATD)}
The term \enquote{technical debt} (TD) was coined by software developer Ward Cunningham. He initially used the metaphor for communicating to non-technical stakeholders the importance of budgeting resources for refactoring~\citep{Cunningham1992TheSystem}. Technical debt, sometimes referred to as tech debt or code debt, is the result of a development team rushing the delivery of projects or features that must later be refactored. In other words, there is a trade-off between delivering high-quality, effective, and optimized code and the short-term benefits of achieving release dates \citep{Brown2010ManagingSystems, Avgeriou2016ManagingEngineering}. A small amount of debt is acceptable because it can increase productivity, but if left uncontrolled, it could end up costing a lot of time, extra work, and money in the future \citep{seaman2011measuring, tom2013exploration}. The \enquote{debt} metaphor further implies that, just as financial debt accrues interest, the longer suboptimal code remains in the system, the more costly and time-consuming it becomes to remove. This emphasizes the compounding nature of technical debt over time, where delayed repayment leads to greater maintenance burdens and reduced productivity.

The concept \enquote{not-quite-right} is commonly used to refer to internal software development activities that are postponed but may lead to future difficulties if not addressed \citep{Brown2010ManagingSystems, Kruchten2012TechnicalPractice}. It represents the debt incurred when development teams opt for quick and easy solutions in the short term, risking long-term negative impacts.

TD encompasses various suboptimal aspects of software that developers acknowledge but lack the immediate resources to address. These may include outdated documentation, missing tests, complex code in need of refactoring, and known defects left unresolved \citep{Zazworka2013ADebt, Spinola2013InvestigatingOpinion}. Such deficiencies can degrade software quality over time, making maintenance more challenging and causing unforeseen delays when implementing necessary changes.

In recent years, numerous empirical studies have focused on identifying technical debt through textual artifacts, particularly source code comments and other developer-written text, a practice commonly referred to as self-admitted technical debt (SATD) \citep{Potdar2014AnDebt}. The primary sources for SATD detection include code comments, commit messages, issue tracker discussions, and other natural language-based software artifacts.

As SATD is expressed in natural language, its detection relies heavily on language processing techniques to analyze and classify technical debt instances. Therefore, SATD detection can be regarded as an application of NLP within the broader context of software engineering, as it relies heavily on natural language techniques to identify and classify SATD.


\subsection{NLP in SATD Detection}
Natural Language Processing (NLP) is an interdisciplinary field that studies the interaction between computers and human language. It includes tasks such as speech recognition, natural language understanding, and natural language generation \citep{liddy2001natural, Chowdhury2003NaturalProcessing}. Within the scope of SATD detection, NLP is applied in a more focused way as it provides the means to transform developer-authored textual artifacts such as code comments, commit messages, issue tracker discussions, and pull requests into structured features that automated models can analyze \citep{haiduc2016use, Li2022AutomaticSources}. For instance, code comments often contain explicit markers of pending work (e.g., TODO) or implicit expressions of design compromise, which can be mined using NLP-based approaches \citep{storey2008todo, maalej2010can, shokripour2013so}.

Unlike traditional static code analysis approaches that detect proxies of technical debt, such as code smells or rule violations, SATD detection aims to uncover technical debt as explicitly acknowledged by developers themselves \citep{Potdar2014AnDebt, Maldonado2017UsingDebt}. This makes SATD a unique and valuable complement to other TD detection methods, as it directly reflects developers' awareness of design trade-offs and postponed tasks \citep{Maldonado2015DetectingDebt}.

In prior studies, SATD detection has been approached from two complementary perspectives. Some research focuses on determining whether a given software artifact contains any indication of debt \citep{Maldonado2017UsingDebt, Dai2017DetectingTrackers}, thereby providing an overview of its extent within a system. Other research takes this further by distinguishing between different forms of debt, such as design, architecture, requirement, or test debt, based on the linguistic patterns found in the text \citep{Maldonado2017UsingDebt, Xavier2020BeyondSystems, Chen2022MulticlassXGBoost}. This finer-grained view is particularly useful because different types of SATD have distinct implications for maintenance and repayment strategies \citep{Sierra2019ADebt}. Together, these perspectives highlight how detection approaches can support both the recognition and the deeper understanding of technical debt in practice \citep{maldonado2017empirical}.

Over time, SATD detection techniques have evolved in line with advances in NLP, moving from simple keyword-based heuristics to classical machine learning classifiers trained on TF-IDF features, and more recently to deep learning and Transformer-based models that leverage semantic embeddings \citep{Maldonado2017UsingDebt, li2023automatic, sutoyo2024deep}. Although these approaches demonstrate significant progress, the field is still maturing and continues to face open questions that require systematic investigation. This underlines the importance of consolidating and critically examining the state of the art, which is the focus of this review.

While the background section introduced SATD and its detection techniques, a systematic investigation is required to consolidate existing research findings. To achieve this, we employ a structured literature review methodology. The following section outlines our research design, including our research questions, study selection process, and data extraction methods, ensuring a rigorous and reproducible review.

\section{Research Methodology}
\label{section_rm}
\begin{figure*}[htpb!]
    \centerline{\includegraphics[trim=5.5cm 4.2cm 5.5cm 4.9cm, clip, width=300px]{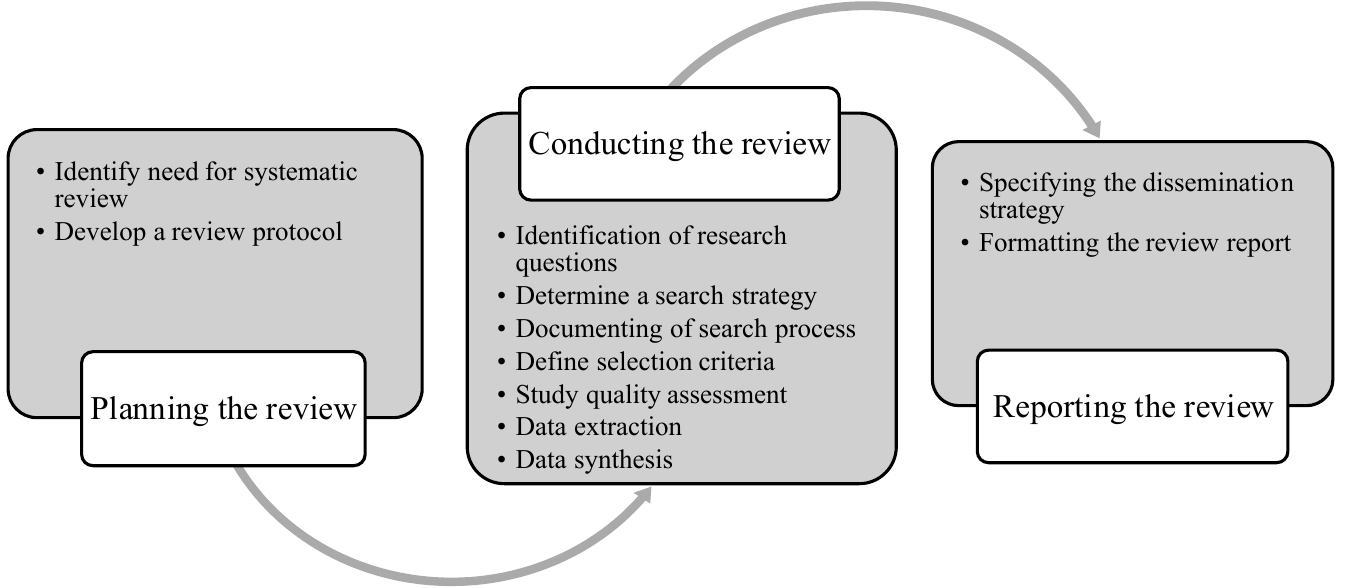}} %
    \caption{Graphical representation of the SLR process used in this article.\label{fig:slr_process}}
\end{figure*}

This study is carried out in accordance with the methodology of conducting an SLR described by Kitchenham et al.~\citep{BarbaraKitchenham2004ProceduresReviews, Kitchenham2006Evidence-basedReviews, Kitchenham2007GuidelinesEngineering}. The procedures for SLR are divided into three stages: planning, conducting, and reporting \citep{Kitchenham2007GuidelinesEngineering}. We developed a review protocol during the planning stage that included the following steps: i) identification of research topics, ii) creation of a search strategy, iii) study selection criteria, iv) study quality assessment, v) data extraction procedure, and vi) data synthesis process (as illustrated in Figure~\ref{fig:slr_process}).


The review process comprises several steps. Initially, research questions were formulated for the SLR. Subsequently, a search strategy was devised, involving the development of search strings and the selection of online databases to locate primary studies. The third stage encompassed the identification of pertinent studies aligned with the research questions, along with the establishment of inclusion and exclusion criteria. To ensure comprehensive coverage, a bidirectional snowballing approach was employed. This involved examining the references cited in the source articles and exploring articles that cited these sources, using Google Scholar\footnote{\url{https://scholar.google.com}} as the primary tool, as recommended in prior guidelines \citep{brereton2007lessons}. Subsequently, we developed a data extraction form to address the research questions. A data extraction form was then crafted to address the research questions. Lastly, a data synthesis approach was developed to analyze and summarize the research findings.


\subsection{Research Questions}
\label{rqs}
Our SLR reviews all relevant research outlining the various SATD detection techniques. To define the search string, we systematically broke down the research questions using the PICOC framework (Population, Intervention, Comparison, Outcome, and Context), as recommended by Petticrew and Roberts \citep{petticrew2008systematic}. Boolean operators were applied to structure the search effectively: OR was used to include synonyms and alternative terms, while AND linked the key terms, as shown in Table~\ref{tab:tb_picoc}. This approach ensured a comprehensive and focused search strategy.

\begin{table}[ht!]
    \centering
    \caption{PICOC Criteria.} 
    \begin{tabular}{l>{}m{7cm}>{}m{8cm}>{}m{6cm}}
        \toprule
        \textbf{Parameter} & \textbf{Description} & \textbf{Keywords Used}  \\
        \midrule
        Population & Field of application & \enquote{self-admitted technical debt} OR SATD\\
        Intervention & Approach or method for dealing with a certain problem & \enquote{nlp} AND \enquote{technical debt}\\
        Comparison & Not applicable & Not applicable\\
        Outcome & Related to factors of importance to practitioners & \enquote{detection} OR \enquote{identification} OR \enquote{prediction}\\
        Context & Context in which intervention take places & \enquote{software development} OR \enquote{software engineering}\\
        \bottomrule
    \end{tabular}
    \label{tab:tb_picoc}
\end{table}


According to the PICOC in Table~\ref{tab:tb_picoc}, the main objective of this SLR is to investigate, identify, and classify all relevant evidence from the primary studies that focus on using NLP to detect SATD. Therefore, this SLR aims to answer the following questions:


\textbf{RQ1: What types of techniques have been employed to detect SATD in software development, and how do these techniques handle different types of textual artifacts?} \\
\textit{Rationale:} This question aims to categorize and analyze the various types of techniques that have been proposed for SATD detection across different types of textual artifacts, providing insight into trends and gaps in the field.



\textbf{RQ2: What levels of automation have been proposed in methodologies for detecting SATD in software development, and to what extent do these methods rely on human interventions across different types of textual artifacts?} \\
\textit{Rationale:} This question focuses on understanding the extent to which SATD detection methods incorporate automation, highlighting the progress and identifying gaps in automating the detection process across different types of textual artifacts. 


\textbf{RQ3: Which SATD detection techniques are reported as the most effective based on evaluation metrics?}\\
\textit{Rationale:} This question aims to review and synthesize the reported precision, recall, and F1 scores of the identified techniques to provide an overview of their observed performance. The goal is not to perform an independent benchmark but to consolidate how techniques have been reported in the literature and to identify which are most frequently reported as achieving the highest effectiveness.

\textbf{RQ4: What tools have been proposed for SATD detection?}\\
\textit{Rationale:} This question aims to evaluate the current landscape of tools designed for detecting SATD from the last decade. 



\textbf{RQ5: What are the primary challenges associated with the detection of SATD?}\\
\textit{Rationale}: This question focuses on uncovering the common obstacles faced when detecting SATD, which could help direct future research and tool development.

\subsection{Search Protocol}

Before initiating the SLR planning stage, it is essential to conduct a pilot search in databases such as Scopus\footnote{\url{https://www.scopus.com}} to determine whether existing SLRs have already addressed the same research theme. If a relevant SLR is found, conducting a new one may not be necessary \citep{popay2005developing}.

Our preliminary search revealed no existing SLRs encompassing the breadth of approaches for detecting SATD. Consequently, this SLR fills a critical gap in the literature and offers a timely contribution to the field.

The SLR's search strategy consists of three stages: pilot search, primary search, and snowballing. In the pilot search, multiple search strings are used in online databases to identify the most effective. The primary search uses the identified search string to retrieve relevant literature from online databases. The search concludes with a snowball search \citep{Wohlin2014GuidelinesEngineering}. The pilot and primary search utilized six online databases: Web of Science, Scopus, IEEE Xplore Digital Library, ScienceDirect, ACM Digital Library, and Springer. These databases, as demonstrated by previous software engineering SLRs \citep{mourao2020performance, abuhassan2021software, garces2021three, melo2022identification}, provide an excellent overview of the software engineering literature and are sufficiently comprehensive to cover our scope. Moreover, we employed the \enquote{Advanced Search} feature in these databases to obtain full-text and metadata results.

The first author communicated the results of a pilot search using the search string S1 to the second author. The second author randomly selected 10\% of the articles to check their relevance. Based on the results, the second author proposed modifying the search string to S2 because the results from S1 remained too general. Similar to the initial process, the iteration was repeated until it was decided to use the final search string, namely S3 (see Table~\ref{tab:tb_search_string}).

\begin{table}[htb!]
    \centering
    \caption{General search strings utilized during pilot search.}
    \begin{tabular}{l>{}m{12.5cm}}
        \toprule
        \textbf{String ID} & \textbf{Search String} \\
        \midrule
        S1 & (\enquote{self-admitted technical debt} OR SATD) AND (NLP OR \enquote{natural language processing})\\
        S2 & (\enquote{self-admitted technical debt} OR SATD) AND (NLP OR \enquote{natural language processing}) AND (detection OR identification)\\
        \cellcolor{gray!40}S3 & \cellcolor{gray!40}(\enquote{self-admitted technical debt} OR SATD) OR (\enquote{technical debt} AND NLP) AND (detect* OR identif* OR predict*) AND (\enquote{software engineering} OR \enquote{software development})\\
        \bottomrule
    \end{tabular}
    \label{tab:tb_search_string}
\end{table}

Through a series of pilot searches with multiple search strings, the most effective one (i.e., S3) was selected. The search string was designed with synonyms, alternative keywords, and the Boolean operators `OR' and `AND' to retrieve relevant studies. Building on the earlier pilot search, we expanded search string S3 by adding three supplementary terms that are commonly used in SATD identification tasks: detect, identify, and predict. Additionally, the wildcard asterisk (*) was used to broaden the search for variations of words \citep{aromataris2014constructing}.

The final search string (i.e., S3) was then applied to the six selected online databases to identify relevant primary studies addressing the research questions described in Subsection~\ref{rqs}. We utilized the \enquote{Advanced Search} feature in these databases to obtain full-text and metadata results, as summarized in Table~\ref{tab:tb_search_result}.

\begin{table}[htb!]
    \centering
    \caption{Search results from selected databases.}
    \begin{tabular}{clr}
        \toprule
        \textbf{No} & \textbf{Database} & \textbf{Search Result}\\
        \midrule
        1 & Web of Science & 56\\
        2 & Scopus  & 1,118\\
        3 & IEEE Xplore & 184\\
        4 & ACM Digital Library & 49\\
        5 & Springer & 181\\
        6 & ScienceDirect & 69\\
        \midrule
        & Total & 1,659\\
        \bottomrule
    \end{tabular}
    \label{tab:tb_search_result}
\end{table}

\subsection{Study Selection}
\label{inclusion-exclusion}
\begin{figure}[htb!]
    \centering
    \includegraphics[trim=0.1cm 0.1cm 0.1cm 0.1cm, clip, width=180px]{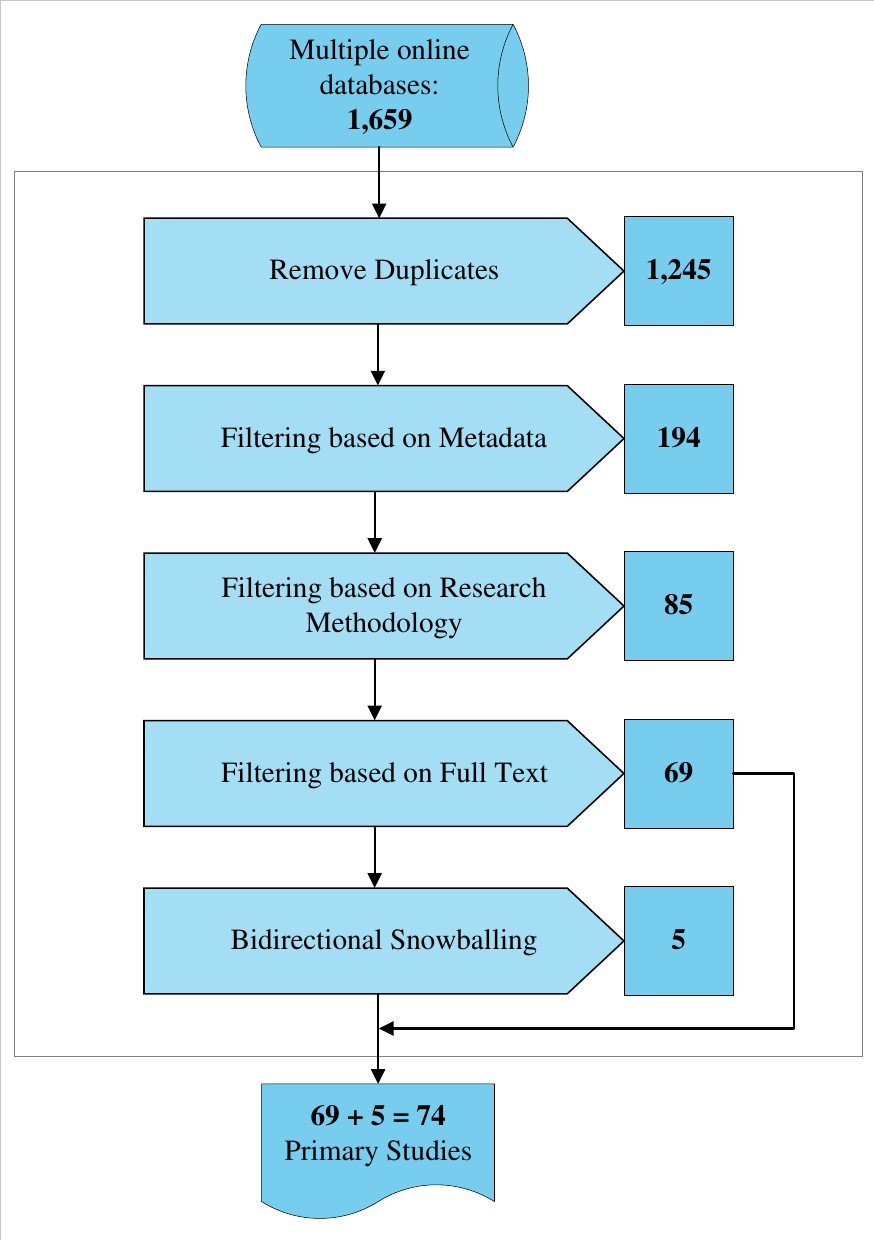}
    \caption{An overview of the steps in determining primary studies.}
    \label{fig:slr_filtering}
\end{figure}

As shown in Figure~\ref{fig:slr_filtering}, we selected primary studies through a staged screening procedure to retain only relevant and sufficiently detailed evidence for this review. The initial search across multiple digital libraries returned 1{,}659 records. After duplicate removal, 1{,}245 unique records remained. We then applied a three-stage screening process guided by the inclusion and exclusion criteria in Table~\ref{tab:tb_inclusion_exclusion}.


First, we performed a metadata screening of titles, abstracts, and keywords to assess topical relevance, which reduced the set to 194 studies. Second, we screened studies for research methodology and eligibility (e.g., whether the study reported an SATD identification approach and provided sufficient methodological detail), yielding 85 studies. Third, we conducted full-text screening to confirm relevance and assess the study's contribution and reporting quality, yielding 69 studies focused on SATD detection. These 69 studies constitute the initial core set of primary studies for our SLR.

\begin{table}[htb!]
    \centering
    \caption{Inclusion and exclusion criteria.}
    \begin{tabular}{l>{}m{15cm}>{}m{13cm}}
        \toprule
        \textbf{No} & \textbf{Inclusion criteria}\\
        \midrule
        1 & The article must be peer-reviewed and published at a conference, workshop, or journal\\
        2 & The article must be accessible online (i.e., articles can be fully accessed and downloaded)\\
        3 & For research that has both publication types (journal and proceeding version), only the journal version will be selected\\
        4 & The study must discuss techniques or approaches for identifying and measuring the performance of SATD detection\\
        5 & The study must have been published between 2014 and April 2025\\
        6 & The study reported must be related to SATD in software engineering or software development\\
        \midrule
        \textbf{No} & \textbf{Exclusion Criteria}\\
        \midrule
        1 & Articles written in languages other than English\\
        2 & Works in the form of dissertations, theses, and books\\
        3 & Articles that are still in the form of proposals\\
        4 & The article does not answer at least one research question\\
        5 & All other secondary and tertiary studies\\

        \bottomrule
    \end{tabular}
    \label{tab:tb_inclusion_exclusion}
\end{table}

To ensure the reliability of our inclusion and exclusion process, we followed the recommendations of Kitchenham and Charters~\citep{Kitchenham2007GuidelinesEngineering}, who suggest that consistency of study selection can be verified through a test-retest procedure as an alternative to inter-rater agreement. The first author conducted the initial screening of studies across all stages (metadata, research methodology, and full-text analysis). To check the consistency of the labeling process, the first author also applied a test-retest approach by re-evaluating a random sample of the studies. In addition, to minimize potential bias, a subset of 15\% of the studies at each filtering stage was independently reviewed by the second author. Both validation steps yielded full agreement with the initial decisions, and no disagreements were found, indicating that the risk of misclassification was negligible. The replication package, including the screening spreadsheet and supplementary material, is publicly available.

After completing these validation steps, both authors agreed to select 69 relevant studies from six online databases for inclusion in the systematic review. This rigorous and transparent process ensured the inclusion of only the most relevant and high-quality studies.

The replication package, including the screening spreadsheet and supplementary material, is publicly available.\footnote{\url{https://github.com/edisutoyo/satd-detection-slr}}

\subsection{Snowballing Process}
To improve the completeness of the review, we employed bidirectional snowball sampling in accordance with Wohlin's guidelines \citep{Wohlin2014GuidelinesEngineering}. In line with the practical recommendations of Brereton et al.~\citep{brereton2007lessons}, we used Google Scholar to support both backward and forward snowballing. Backward snowballing was performed by manually screening the reference lists of the included studies, whereas forward snowballing was conducted by identifying and screening papers that cited those included studies.

For backward snowballing, references within the files of the selected studies were carefully examined. Each reference was assessed based on its title and abstract to determine its relevance. When necessary, the full texts of the referenced studies were retrieved and reviewed for further evaluation. Initially, the collected references were filtered by their titles and abstracts to assess alignment with the research questions. Subsequently, the full texts of potentially relevant studies were analyzed to determine whether they met the inclusion criteria.

To ensure the reliability of the snowballing process, both authors were involved in assessing the relevance of the identified studies. The first author initially screened the references and citations, and the second author independently reviewed the candidate papers and cross-checked their eligibility. Any disagreements were discussed until a consensus was reached. Only one candidate paper required clarification at this stage, which was resolved through discussion. This indicates that disagreements were minimal and resolvable.

This rigorous process helped uncover additional studies that might have been missed during the initial search. Five additional studies were included as a result. In the final stage, the 69 studies selected through full-text screening were combined with the five studies identified through bidirectional snowballing, yielding a total of 74 primary studies. This collaborative effort ensured that the selection process was both thorough and objective, thereby leading to a comprehensive review of the existing literature on SATD detection approaches.


\subsection{Quality Assessment Criteria}
After finalizing the list of articles for the systematic literature review (SLR), we performed a quality assessment to ensure that the selected papers provided the necessary information to answer our research questions. We assessed the quality of the selected articles using the questions in Table \ref{tab:tb_quality_assessment}, following the procedure recommended by Kitchenham \citep{BarbaraKitchenham2004ProceduresReviews}. All 74 articles deemed relevant satisfied the predefined quality assessment criteria. Accordingly, our SLR is based on these 74 primary studies.

\begin{table}[htb!]
    \centering
    \caption{Quality assessment criteria.}
    \begin{tabular}{l>{}m{13.7cm}>{}m{12cm}}
        \toprule
        \textbf{No} & \textbf{Questions}\\
        \midrule
        1 & Are the research objectives or research questions in the article clear?\\
        2 & Does the study include reviewing the relevant literature or background?\\
        3 & Does the paper describe a research methodology?\\
        4 & Does the article employ an algorithm or technique for SATD detection?\\
        5 & Does the study provide research findings and answer the research questions?\\
        6 & Does the article report metric evaluation performance, or at least report results of the approach?\\
        7 & Does the article clearly state the dataset used?\\
        \bottomrule
    \end{tabular}
    \label{tab:tb_quality_assessment}
\end{table}

All of the listed articles were published between January 2014 and April 2025. Our searches for articles were completed by early May 2025 and were supplemented by articles identified through bidirectional snowballing. As a result, it is highly unlikely that we missed any published articles during our period due to publication delays.

\subsection{Data Extraction}
\label{data-extraction}
We selected the final relevant articles after applying the inclusion and exclusion criteria and conducting the quality evaluation. We determined the required information and attributes for data summarization to address our research questions. This facilitated data extraction, in which we read the full text of each article and extracted the relevant information based on specified attributes. To answer the research questions mentioned in Subsection \ref{rqs}, we extracted data items listed in Table~\ref{tab:tb_data_extraction} from each selected study. The extracted data was then entered into a spreadsheet for further analysis.

As part of the extraction, we also examined the evaluation setup adopted by each study. Since our review specifically focuses on within-project evaluations (e.g., k-fold cross-validation or random splits performed on the same dataset or project), we excluded cross-project or out-of-distribution evaluations from the performance synthesis in RQ3. This decision was made to ensure comparability of results across studies and to avoid conflating performance metrics obtained under substantially different evaluation settings.

\begin{table}[htb!]
    \centering
    \caption{Data extracted from each study.}
    \begin{tabular}{l>{}m{3.2cm}>{}m{9.9cm}>{}m{2.5cm}}
        \toprule
        \textbf{No} & \textbf{Data Name} & \textbf{Description} & \textbf{Relevant RQ}\\
        \midrule
        1 & Title & Title of the study & Overview\\
        2 & Author & Name of the authors & Overview\\
        3 & Year & The publication year of the study & Overview\\
        4 & Venue & The name of the publication venue of the study & Overview\\
        5 & Publication type & Journal, conference, workshop, or book chapter & Overview\\
        6 & Algorithms used & Algorithm or model used in the study & RQ1, RQ2, RQ4\\
        7 & Tool proposed & Name or description of any tool proposed in the study & RQ1, RQ2, RQ4\\
        8 & TD types & The types of TD, e.g., design or requirement debt & RQ2, RQ3\\
        9 & Number of class & Number of class(es)/TD types used & RQ2, RQ3\\
        10 & Textual artifacts used & Data source or software artifacts used & RQ1, RQ4\\
        11 & Metrics evaluation used & Types of metric evaluations used for performance measurement & RQ3\\
        12 & Performance summary & All included studies report within-project validation (e.g., k-fold CV, random splits). Cross-project results are not synthesized in this review & RQ3\\
        13 & Challenges & Issues or challenges related to SATD detection & RQ5\\
        \bottomrule
    \end{tabular}
    \label{tab:tb_data_extraction}
\end{table}

To ensure the reliability of the data extraction and categorization process, we followed the guidance of Kitchenham and Charters~\citep{Kitchenham2007GuidelinesEngineering}. As a `\textit{gold-standard procedure}', they recommend that data extraction should be performed independently by at least two reviewers whenever feasible. Because full dual independent extraction for all included studies was not feasible under our resource constraints, we applied two complementary reliability checks: (i) a test--retest procedure in which the first author re-extracted and re-categorized a random sample after an interval to assess decision stability, and (ii) an independent cross-check by the second author on 15\% of the extracted items.\footnote{This conforms to the guidelines as suggested in section 6.4.3 of ~\citep{Kitchenham2007GuidelinesEngineering}: `\textit{Alternatively, a test-retest process can be used where the researcher performs a second extraction from a random selection of primary studies to check data extraction consistency'}. The shared spreadsheet used to take these decisions is available in the supplementary material at https://github.com/edisutoyo/satd-detection-slr)} These checks are in line with the guidelines of Kitchenham and Charters~\citep{Kitchenham2007GuidelinesEngineering}, and resulted in full agreement with the initial decisions, and no disagreements were observed. For transparency and replicability, the complete dataset and categorization results are made available in our replication package.



With the methodology in place, we present our findings from the 74 selected studies. The results are organized according to our research questions, beginning with an analysis of the techniques used for SATD detection, followed by an evaluation of automation levels, effectiveness metrics, available tools, and key challenges in this domain.

\section{Results}
\label{section_results}
This section addresses our research questions and discusses the findings from the 74 primary studies included in this review.

\subsection{Results - RQ1: Techniques employed for detecting SATD in software development across textual artifacts}
\label{rq-techniques}
\begin{figure}[htpb!]
    \centering
    \includegraphics[width=340px]{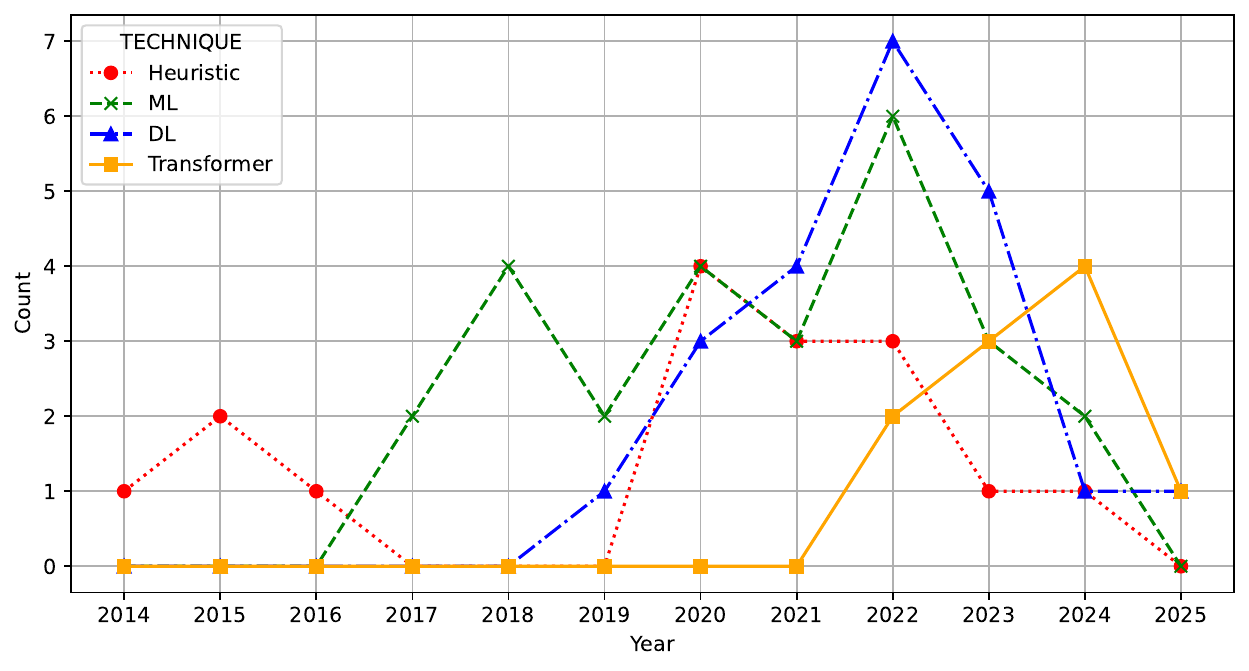}
    \caption{Distribution of articles by technique over the years.}
    \label{fig:chart-technique}
\end{figure}

To classify the relevant studies, we followed a bottom-up categorization process. During data extraction, we recorded the algorithms and models used in each primary study (see Table~\ref{tab:tb_data_extraction}) and grouped them into higher-level categories based on commonalities observed in the literature and prior surveys. This process led to four main categories (see Table~\ref{tab:tb_approach} and Figure~\ref{fig:chart-technique}): (i) heuristic-based approaches (rule-based, keyword, or patterns matching methods), (ii) machine learning approaches (classical supervised models such as SVM, NB, and RF), (iii) deep learning approaches (neural architectures such as CNN, LSTM, BiLSTM, and GNN), and (iv) Transformer-based approaches. While Transformers are technically a specialized form of deep learning, we treat them as a distinct category because they represent a clear methodological shift: unlike recurrent models such as LSTM, Transformers rely on self-attention mechanisms \citep{vaswani2017attention}, are more scalable, and have become the dominant paradigm in recent SATD studies. This separation follows trends observed in both NLP and SE literature \citep{gruetzemacher2022deep, hou2024large}, where Transformers are increasingly treated as a distinct family of models due to their state-of-the-art performance and widespread adoption. Moreover, recent large language models (LLMs) are built on the Transformer architecture \citep{wang2019language}, further reinforcing the distinctiveness of this category.


\begin{table}[htpb!]
    \centering
    \caption{Technique identified. Studies that fall into more than one technique are highlighted with square brackets.} 
    \begin{tabular}{l>{}m{6.5cm}>{}m{7cm}}
        \toprule
        \textbf{Technique} & \textbf{Paper Reference} & \textbf{Technique examples}\\
        \midrule
        Heuristic & No. \citep{Potdar2014AnDebt, Farias2015AComments, Maldonado2015DetectingDebt, Bavota2016ADebt, Li2020IdentificationTrackers, Farias2020IdentifyingVocabulary, Al-Slais2021TowardsComments, Maipradit2020WaitDebt, Rajalakshmi2021ClassificationAnalytics, Guo2021HowStudy, Azuma2022AnDockerfile, Russo2022WeakSATD:Debt, sheikhaei2023automated, farias2021comment, qu2022we, ishimoto2024empirical}
        & Annotation tags, Textual patterns matching, etc.
        \\

        \midrule
        Machine learning & No. \citep{Dai2017DetectingTrackers, Maldonado2017UsingDebt, Flisar2018EnhancedIdentification, Wattanakriengkrai2019IdentifyingIDF, Huang2018IdentifyingMining, Liu2018SATDTool, Wattanakriengkrai2019AutomaticIDF, Flisar2019IdentificationEmbedding, Maipradit2020AutomatedDebt, Xavier2020BeyondSystems, Rantala2020PredictingSelection, Rantala2020PrevalenceKL-SATD, Santos2020Self-AdmittedNetwork, Sridharan2021DataDetection, Sala2021DebtHunter:Debt}

        \citep{Yu2020IdentifyingApproach, Chen2022MulticlassXGBoost, ALOMAR2022102693, yin2023two, satyaAutomated, pinna2023investigation, gama2024towards, phaithoon2021fixme, kaur2024embracing, shahzeidi2025hybrid}\textsuperscript{,[}\citep{Sharma2022Self-admittedCauses, Khan2022AutomaticPackages, Sabbah2023Self-admittedEmbeddings, li2023automatic}\textsuperscript{]}
        & NB, ME, SVM, kNN, etc.\\
        \midrule   

        Deep learning & No. \citep{Ren2019NeuralExplainability, Wang2020DetectingNetworks, Santos2020LongDetection, Zhu2021DetectingCNN-BiLSTM, Santos2021EvaluatingComments, Yu2021UsingDebt, Yu2021UsingDebt, Zhuang2022AnDetection, Xiao2022CharacterizingSystems, Yin2022DeepDebt, YU2022111219, Li2022IdentifyingLearning, DiSalle2022PILOT:Debt, Li2022Self-admittedNetworks, shahzeidi2025hybrid}
        
        \citep{li2023debtviz, qu2023deep, yu2023detecting, aiken2023measuring, Zhu2023SCGRU:Oversampling}\textsuperscript{,[}\citep{Sharma2022Self-admittedCauses, Khan2022AutomaticPackages, Sabbah2023Self-admittedEmbeddings, li2023automatic, sutoyo2024deep, sheikhaei2024empirical, shahzeidi2025hybrid}\textsuperscript{]}
        & CNN, LSTM, BiLSTM, etc.\\
        \midrule
        Transformer & No. \citep{skryseth2023technical, gong2023identifying, qu2022empirical, karmakar2022experience, li2024large, gu2024self, sheikhaei2024empirical, shivashankar2025beacon}\textsuperscript{,[}\citep{Sharma2022Self-admittedCauses, Khan2022AutomaticPackages, Sabbah2023Self-admittedEmbeddings, sutoyo2024deep}\textsuperscript{]}
        & BERT, ALBERT, ChatGPT, etc.\\
        
        \bottomrule
    \end{tabular}
    \label{tab:tb_approach}
\end{table}

The categorization process followed the reliability procedures described in Subsection~\ref{data-extraction}, which included test-retest and cross-checking by the second author. Only one disagreement was recorded and resolved through consensus, indicating that the grouping was stable and consistent.

Figure~\ref{fig:chart-technique} presents the distribution of published papers categorized by their detection techniques, including heuristic-based methods, ML, DL, and Transformer-based approaches. Over the past decade, various approaches to SATD detection have emerged, evolving from heuristic-based methods to more advanced ML and DL techniques. Early methods relied primarily on traditional heuristic techniques, focusing on keyword detection and textual patterns in code comments. As depicted in Figure~\ref{fig:chart-technique}, heuristic methods initially dominated, with limited exploration of advanced computational methods. From 2017 to 2019, ML techniques were increasingly introduced to improve SATD detection accuracy. Between 2020 and 2022, DL and Transformer architectures surged, marking a paradigm shift toward data-driven techniques capable of capturing more complex semantic patterns. From 2023 onward, DL and Transformer models became the predominant approaches, with Transformers slightly surpassing DL in 2024 and maintaining their momentum into early 2025. This trend highlights a growing reliance on Transformer-based models for advancing SATD detection.

Heuristic-based techniques have long played a fundamental role in detecting SATD. These approaches typically rely on lightweight, rule-driven text processing and pattern matching rather than model training. As summarized in Table~\ref{tab:tb_approach}, heuristic pipelines may also incorporate manual inspection and Parts-of-Speech (POS) tagging \citep{Farias2015AComments, Farias2020IdentifyingVocabulary, Al-Slais2021TowardsComments, qu2022we} to support the categorization of debt-related expressions. In addition, regular expressions \citep{Bavota2016ADebt, Russo2022WeakSATD:Debt}, keyword-labeled patterns \citep{Rantala2020PrevalenceKL-SATD}, and other textual pattern rules \citep{Azuma2022AnDockerfile} are widely used to semi-automate the detection of recurring SATD indicators in large codebases.

As software systems have grown in complexity, the need for more advanced technical debt detection methods has intensified, driving the adoption of ML classifiers to enhance accuracy and scalability. These algorithms include Na\"ive Bayes (NB) \citep{Dai2017DetectingTrackers, Liu2018SATDTool, Xavier2020BeyondSystems}, Maximum Entropy (ME) \citep{Maldonado2017UsingDebt, ALOMAR2022102693}, Support Vector Machines (SVM) \citep{Flisar2018EnhancedIdentification}, and k-Nearest Neighbor (kNN). Logistic Regression (LR) \citep{Tu2022DebtFree:Learning} and Random Forest (RF) \citep{Wattanakriengkrai2019AutomaticIDF, Rajalakshmi2021ClassificationAnalytics} are also employed, with various degrees of accuracy and interpretability in detecting patterns linked with technical debt. Furthermore, XGBoost \citep{Chen2022MulticlassXGBoost, yin2023two} and LightGBM~\citep{yin2023two}, relatively new additions to the ML classifier, improve performance by refining gradient boosting approaches. 

The rise of DL has introduced even more advanced techniques for detecting SATD. Convolutional Neural Network (CNN) \citep{Ren2019NeuralExplainability, Zhuang2022AnDetection, Yin2022DeepDebt, li2023debtviz}, Long Short-Term Memory (LSTM) \citep{Santos2020LongDetection, Santos2021EvaluatingComments}, and Bidirectional Long Short-Term Memory (BiLSTM) \citep{Wang2020DetectingNetworks, Yu2021UsingDebt} are among the most prominent techniques used. They are particularly useful in processing structured data and identifying patterns that are not easily captured by traditional ML algorithms. DL techniques have been extensively studied and have been shown to be more effective for automatically detecting technical debt.

The transformer is the most recent development in SATD detection. By offering excellent context understanding and language representation, these models, namely Bidirectional Encoder Representations from Transformers (BERT) \citep{skryseth2023technical, gong2023identifying, gu2024self, qu2022empirical, sutoyo2024deep}, A Lite BERT (ALBERT) \citep{karmakar2022experience}, Robustly optimized BERT approach (RoBERTa) \citep{karmakar2022experience}, DistilBERT \citep{karmakar2022experience}, DistilRoBERTa~\citep{shivashankar2025beacon}, as well as more recent large-scale models such as Flan-T5 \citep{sheikhaei2024empirical} and ChatGPT \citep{li2024large}, have transformed the field of NLP. Transformer excels at handling large-scale data and understanding the nuances in code comments and other textual software artifacts that signify technical debt. The literature reflects a growing trend toward using these models to achieve higher accuracy and more robust identification processes, particularly for large, complex datasets typical of modern software systems.

\begin{table}[htpb!]
    \centering
    \caption{Textual artifacts used. Studies that use more than one textual software artifact are highlighted with square brackets.}
    \begin{tabular}{m{1.8cm}>{}m{4.8cm}|>{}m{2cm}>{}m{2.3cm}>{}m{2.3cm}>{}m{2cm}}
        \toprule
        \textbf{Artifact} & \textbf{Paper Reference} & \textbf{Heuristic} & \textbf{ML} & \textbf{DL} & \textbf{Transformer} \\
        \midrule
        Code comment & No. \citep{Potdar2014AnDebt, Farias2015AComments, Maldonado2015DetectingDebt, Bavota2016ADebt, Maldonado2017UsingDebt, Flisar2018EnhancedIdentification, Wattanakriengkrai2019IdentifyingIDF, Huang2018IdentifyingMining, Liu2018SATDTool, Wattanakriengkrai2019AutomaticIDF, Flisar2019IdentificationEmbedding, Ren2019NeuralExplainability, Maipradit2020AutomatedDebt}\textsuperscript{,} 
        
        \citep{Wang2020DetectingNetworks, Farias2020IdentifyingVocabulary, Farias2020IdentifyingVocabulary, Rantala2020PredictingSelection, Rantala2020PrevalenceKL-SATD, Santos2020Self-AdmittedNetwork, Al-Slais2021TowardsComments, Maipradit2020WaitDebt, Sridharan2021DataDetection, Sala2021DebtHunter:Debt, Zhu2021DetectingCNN-BiLSTM, Santos2021EvaluatingComments, Guo2021HowStudy, Yu2021UsingDebt}\textsuperscript{,}
        
        \citep{Zhuang2022AnDetection, Azuma2022AnDockerfile, Xiao2022CharacterizingSystems, Tu2022DebtFree:Learning, Yin2022DeepDebt, YU2022111219, Yu2020IdentifyingApproach, Chen2022MulticlassXGBoost, DiSalle2022PILOT:Debt, ALOMAR2022102693, Li2022Self-admittedNetworks, Sharma2022Self-admittedCauses, Russo2022WeakSATD:Debt}\textsuperscript{,}
        
        \citep{yin2023two, qu2023deep, yu2023detecting, pinna2023investigation, aiken2023measuring, Zhu2023SCGRU:Oversampling, Sabbah2023Self-admittedEmbeddings, gong2023identifying, li2024large, gama2024towards, qu2022we, kaur2024embracing, sheikhaei2024empirical}\textsuperscript{,}
        
        \citep{farias2021comment, phaithoon2021fixme, qu2022empirical, shahzeidi2025hybrid}\textsuperscript{,[}\citep{satyaAutomated, li2023automatic, li2023debtviz, gu2024self, sutoyo2024deep}\textsuperscript{]} 
        & \citep{Potdar2014AnDebt,
        Farias2015AComments,
        Maldonado2015DetectingDebt,
        Bavota2016ADebt,
        Farias2020IdentifyingVocabulary,
        Al-Slais2021TowardsComments,
        Guo2021HowStudy}\textsuperscript{,}
        
        \citep{Azuma2022AnDockerfile,
        Russo2022WeakSATD:Debt,
        qu2022we,
        farias2021comment}
        
        & \citep{Maldonado2017UsingDebt,
        Flisar2018EnhancedIdentification,
        Wattanakriengkrai2019IdentifyingIDF,
        Huang2018IdentifyingMining,
        Liu2018SATDTool,
        Wattanakriengkrai2019AutomaticIDF,
        Flisar2019IdentificationEmbedding}\textsuperscript{,}
        
        \citep{Maipradit2020AutomatedDebt,
        Rantala2020PredictingSelection,
        Rantala2020PrevalenceKL-SATD,
        Maipradit2020WaitDebt,
        Sala2021DebtHunter:Debt,
        Xiao2022CharacterizingSystems,
        Tu2022DebtFree:Learning}\textsuperscript{,}
        
        \citep{Yu2020IdentifyingApproach,
        Chen2022MulticlassXGBoost,
        ALOMAR2022102693,
        Sharma2022Self-admittedCauses,
        yin2023two}\textsuperscript{,[}\citep{satyaAutomated}\textsuperscript{]}\textsuperscript{,}
        
        \citep{pinna2023investigation}\textsuperscript{,}       
        \citep{gama2024towards,
        phaithoon2021fixme, kaur2024embracing} 
        & \citep{Ren2019NeuralExplainability,
        Wang2020DetectingNetworks,
        Santos2020LongDetection,
        Santos2020Self-AdmittedNetwork,
        Sridharan2021DataDetection,
        Zhu2021DetectingCNN-BiLSTM,
        Santos2021EvaluatingComments}\textsuperscript{,}
        
        \citep{Yu2021UsingDebt,
        Zhuang2022AnDetection,
        YU2022111219,
        DiSalle2022PILOT:Debt,
        Li2022Self-admittedNetworks}\textsuperscript{,[}\citep{li2023automatic}\textsuperscript{],}
        \textsuperscript{[}\citep{li2023debtviz, sutoyo2024deep}\textsuperscript{],}
        \citep{qu2023deep,
        yu2023detecting,
        Zhu2023SCGRU:Oversampling}\textsuperscript{,}
        
        \citep{gong2023identifying, sheikhaei2024empirical, shahzeidi2025hybrid}
        & \citep{aiken2023measuring,
        Sabbah2023Self-admittedEmbeddings,
        li2024large}\textsuperscript{,[}\citep{gu2024self, sutoyo2024deep}\textsuperscript{],}
        
        \citep{qu2022empirical, sheikhaei2024empirical, shahzeidi2025hybrid}\\

        \midrule


        Issue & No. \citep{Dai2017DetectingTrackers, Xavier2020BeyondSystems, Li2020IdentificationTrackers, Rajalakshmi2021ClassificationAnalytics, Codabux2021TechnicalStudy, Khan2022AutomaticPackages, Li2022IdentifyingLearning, skryseth2023technical, shivashankar2025beacon}\textsuperscript{,}
        
        \textsuperscript{[}\citep{satyaAutomated, li2023automatic, li2023debtviz, gu2024self, sutoyo2024deep}\textsuperscript{]} 
        & \citep{
        Xavier2020BeyondSystems,
        Li2020IdentificationTrackers,
        Codabux2021TechnicalStudy,
        Azuma2022AnDockerfile}
        & \citep{Dai2017DetectingTrackers,
        Xiao2022CharacterizingSystems}\textsuperscript{,[}\citep{satyaAutomated}\textsuperscript{]}
        & \citep{Khan2022AutomaticPackages,
        Li2022IdentifyingLearning}\textsuperscript{,[}\citep{li2023automatic, li2023debtviz, sutoyo2024deep}\textsuperscript{]}
        & \citep{skryseth2023technical, shivashankar2025beacon}\textsuperscript{,[}\citep{gu2024self, sutoyo2024deep}\textsuperscript{]}\\
        \midrule   

        Pull request & No. \citep{karmakar2022experience}\textsuperscript{,[}\citep{satyaAutomated, li2023automatic, gu2024self, sutoyo2024deep}\textsuperscript{]}
         & - & \textsuperscript{[}\citep{satyaAutomated}\textsuperscript{]} & \textsuperscript{[}\citep{li2023automatic, sutoyo2024deep}\textsuperscript{]} & \citep{karmakar2022experience}\textsuperscript{,[}\citep{gu2024self, sutoyo2024deep}\textsuperscript{]}\\

        \midrule   

        Commit message & No. 
        \citep{sheikhaei2023automated, ishimoto2024empirical}\textsuperscript{,[}\citep{satyaAutomated, li2023automatic, gu2024self, sutoyo2024deep}\textsuperscript{]} 
        & \citep{sheikhaei2023automated, ishimoto2024empirical} & \textsuperscript{[}\citep{satyaAutomated}\textsuperscript{]} & \textsuperscript{[}\citep{li2023automatic, sutoyo2024deep}\textsuperscript{]} & \textsuperscript{[}\citep{gu2024self, sutoyo2024deep}\textsuperscript{]}\\
        
        \bottomrule
    \end{tabular}
    \label{tab:tb_dataset}
\end{table}

Building on the approach categorization, it is essential to understand the sources or types of textual artifacts analyzed in the context of SATD. Table~\ref{tab:tb_dataset} categorizes the various studies on SATD detection based on the types of textual artifacts utilized. Code comments (such as from Java projects \citep{Maldonado2017UsingDebt, Ren2019NeuralExplainability}, blockchain projects \citep{qu2022we, pinna2023investigation}, Dockerfile \citep{Azuma2022AnDockerfile, gu2024self}, XML files \citep{Xiao2022CharacterizingSystems, gu2024self}, and quantum software projects \citep{ishimoto2024empirical}), issues (from issue tracking systems \citep{li2023automatic, sutoyo2024deep} and GitHub issue reports \citep{skryseth2023technical, shivashankar2025beacon}), pull requests, and commit messages contain valuable information for SATD detection, providing varying levels of granularity and relevance depending on the artifact analyzed. The application of heuristic, ML, DL, and Transformer-based models across these artifacts shows how approaches have evolved to handle different levels of complexity and scale.

Code comments are the most commonly analyzed artifact for SATD detection, with heuristic, ML, DL, and Transformer-based models being widely employed (e.g., studies \citep{Potdar2014AnDebt, Farias2015AComments, aiken2023measuring}). While issues, pull requests, and commit messages are also explored, they are less frequently studied. Early research on SATD primarily relied on heuristic-based techniques to extract patterns from developer language in code comments and issues. However, more recent studies have incorporated ML, DL, and Transformer models (e.g., \citep{li2023automatic, aiken2023measuring, karmakar2022experience, shivashankar2025beacon, li2024large}) for a deeper and more automated analysis, especially when dealing with larger and more complex datasets. The increasing adoption of Transformer models reflects the field's shift towards more sophisticated and scalable SATD detection across various textual artifacts.

Meanwhile, there has been a noticeable shift towards using a more diverse set of textual artifacts for detecting SATD. Furthermore, studies that utilize more than one textual software artifact (e.g., code comments combined with issue trackers or commit messages) are highlighted with square brackets. The increased use of pull requests and commit messages (e.g., \citep{satyaAutomated, li2023automatic, gu2024self, sutoyo2024deep}) as sources for SATD detection in recent studies signals a growing recognition of their value in providing temporal and contextual data on technical debt. As the field advances, there is an increasing emphasis on automation and on leveraging diverse textual artifacts to improve the scalability and reliability of SATD detection across varied software projects. The studies that employed multi-artifacts represent an effort to enhance the accuracy and robustness of SATD detection by cross-referencing different sources of developer commentary and interaction.

Our synthesis reveals a clear research bias, with most SATD detection studies focusing heavily on code comments, whereas other artifacts, such as issue trackers, commit messages, and pull requests, remain underexplored. This concentration on comments is understandable given their accessibility, yet it limits the diversity and generalizability of current approaches. Quantifying this imbalance shows that although research has advanced techniques (from heuristic to Transformer models), it has largely done so within a narrow scope of artifacts. Highlighting this gap provides guidance for future work to expand the examination of underutilized artifacts, thereby enabling a more comprehensive understanding of SATD.

\begin{tcolorbox}[colback=gray!5!white, colframe=black, title=Answer to RQ1, boxrule=0.5pt]
Over time, SATD detection has evolved from heuristic methods to ML and DL, with Transformer now leading due to its strength in capturing complex textual dependencies. The use of multiple artifacts has increased, yet most studies still focus on code comments, leaving issues, pull requests, and commit messages underrepresented. Quantifying this imbalance underscores the need for future research to broaden its focus toward diverse artifacts.
\end{tcolorbox}


\subsection{Results - RQ2: Proposed methodologies for detecting SATD in software development}
\label{rq-methodologies}
To address this, we categorized the methodologies into three main types: manual, semi-automated, and automated. These categories represent the evolution in improving the accuracy, efficiency, and scalability of SATD detection. Manual methods depend on human expertise to review text artifacts, while semi-automated techniques combine human input with tools to streamline the process. Fully automated methods use advanced algorithms to detect SATD with minimal human intervention. Each approach has its strengths and challenges, which we will explore. 

To ensure the reliability of this categorization, we followed the procedures described in Subsection~\ref{data-extraction}, which included test-retest and cross-checking by the second author. No disagreements were recorded during this process, indicating that the grouping was consistent and reliable.

\begin{figure}[htb!]
    \centering
    \includegraphics[width=340px]{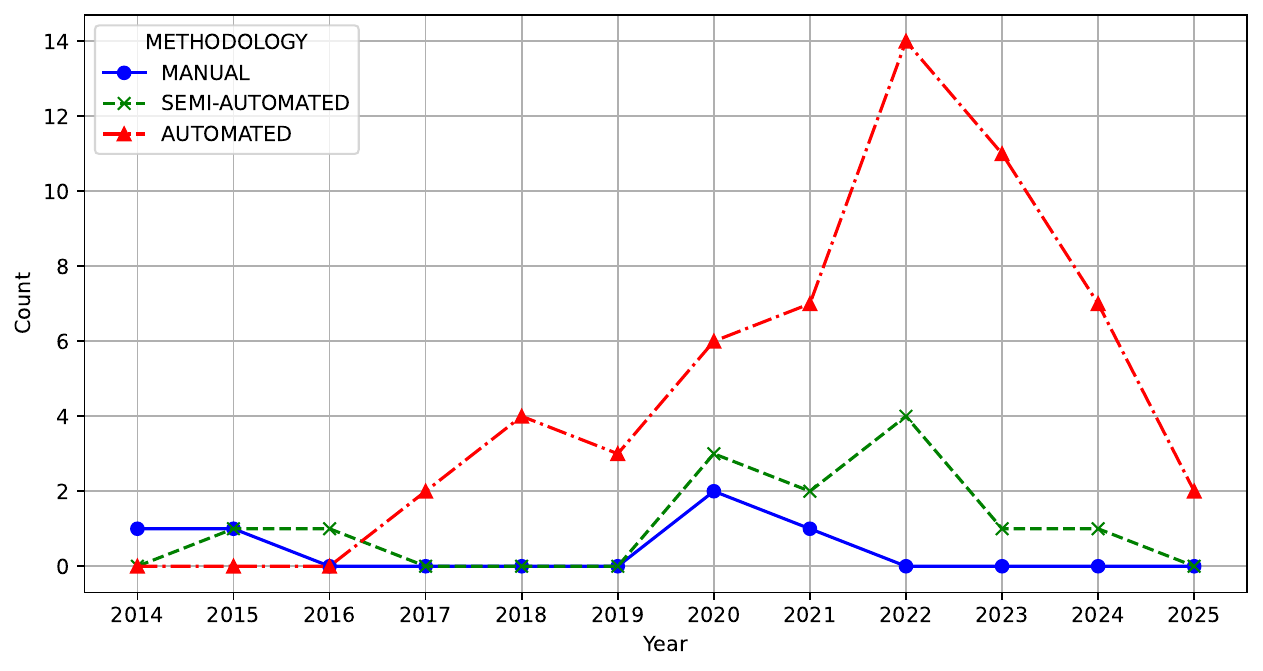}
    \caption{Distribution of articles by methodology over the years.}
    \label{fig:chart-auto}
\end{figure}

Over the past decade, methods for detecting SATD have changed significantly, as shown in Figure~\ref{fig:chart-auto}. There has been a clear shift from manual to more automated techniques, driven by a growing need for efficiency and accuracy in managing technical debt in software development.

From 2014 to 2016, SATD detection mainly relied on manual analyses. This process involved developers or analysts closely examining codebases and documentation to spot technical debt. While these manual methods provided valuable insights, they were time-consuming and prone to human error, underscoring the need for more efficient solutions.

Between 2017 and 2019, the field began transitioning to semi-automated techniques. These methods combined manual efforts with automated tools to enhance the SATD detection process. The figure illustrates a gradual increase in the adoption of semi-automated approaches during this time, often utilizing regular expressions or Part-of-Speech (POS) tagging to reduce developer workload and improve consistency.

The period from 2020 to 2022 marked a significant shift toward fully automated SATD detection methods. As shown in the figure, there has been a notable increase in the use of automated techniques, reflecting the maturation of the field. These approaches employ advanced technologies such as ML and DL to automatically detect technical debt with minimal human intervention.

From 2023 onward, automated methods have become the dominant approach, clearly surpassing manual and semi-automated techniques. In both 2024 and early 2025, automated detection continued to expand, confirming its position as the leading methodology. This trend suggests that the future of SATD detection will likely focus on refining automated tools, with greater reliance on AI-driven techniques to enhance accuracy, efficiency, and adaptability.

As presented in Table~\ref{tab:tb_methodology}, the methodologies are classified into three broad categories: Manual, Semi-Automated, and Automated approaches. These categories are determined by the extent of human intervention and the use of Heuristic, ML, DL, and Transformer models.

\begin{table}[htpb!]
    \centering
    \caption{Methodologies identified.} 
    \begin{tabular}{l>{}m{3.8cm}|>{}m{2.1cm}>{}m{2.2cm}>{}m{2.2cm}>{}m{2.7cm}}
        \toprule
        \textbf{Methodology} & \textbf{Paper Reference} & \textbf{Heuristic} & \textbf{ML} & \textbf{DL} & \textbf{Transformer} \\
        \midrule
        Manual & No. \citep{Potdar2014AnDebt, Maldonado2015DetectingDebt, Xavier2020BeyondSystems, Li2020IdentificationTrackers, Codabux2021TechnicalStudy, ishimoto2024empirical}
         & No. \citep{Potdar2014AnDebt, Maldonado2015DetectingDebt, Xavier2020BeyondSystems, Li2020IdentificationTrackers, Codabux2021TechnicalStudy, ishimoto2024empirical} & - & - & -\\
        \midrule
        Semi-Automated & No. \citep{Farias2015AComments,
        Bavota2016ADebt,
        Farias2020IdentifyingVocabulary,
        Rantala2020PrevalenceKL-SATD,
        Al-Slais2021TowardsComments,
        Guo2021HowStudy,
        Azuma2022AnDockerfile,
        Yu2020IdentifyingApproach,
        Russo2022WeakSATD:Debt,
        sheikhaei2023automated}\textsuperscript{,} 
        
        \citep{qu2022we,
        farias2021comment}
         & No. \citep{
        Farias2015AComments,
        Bavota2016ADebt,
        Farias2020IdentifyingVocabulary,
        Al-Slais2021TowardsComments,
        Guo2021HowStudy}\textsuperscript{,}
        
        \citep{Azuma2022AnDockerfile,
        Russo2022WeakSATD:Debt,
        sheikhaei2023automated,
        qu2022we,
        farias2021comment} & No. \citep{Yu2020IdentifyingApproach, Rantala2020PrevalenceKL-SATD} & - & - \\
        \midrule   
        Automated & No. 
        \citep{Dai2017DetectingTrackers, Maldonado2017UsingDebt, Flisar2018EnhancedIdentification, Wattanakriengkrai2019IdentifyingIDF, Huang2018IdentifyingMining, Liu2018SATDTool, Wattanakriengkrai2019AutomaticIDF, Flisar2019IdentificationEmbedding, Ren2019NeuralExplainability, Maipradit2020AutomatedDebt}\textsuperscript{,} 
        
        \citep{Santos2020LongDetection, Rantala2020PredictingSelection, Santos2020Self-AdmittedNetwork, Maipradit2020WaitDebt, Rajalakshmi2021ClassificationAnalytics, Sridharan2021DataDetection, Sala2021DebtHunter:Debt, Zhu2021DetectingCNN-BiLSTM, Santos2021EvaluatingComments, Yu2021UsingDebt}\textsuperscript{,}
        
        \citep{Zhuang2022AnDetection, Khan2022AutomaticPackages, Xiao2022CharacterizingSystems, Tu2022DebtFree:Learning, Yin2022DeepDebt, YU2022111219, Li2022IdentifyingLearning, Chen2022MulticlassXGBoost, DiSalle2022PILOT:Debt, ALOMAR2022102693, Li2022Self-admittedNetworks}\textsuperscript{,}
        
        \citep{Sharma2022Self-admittedCauses, yin2023two, satyaAutomated, li2023automatic, li2023debtviz, qu2023deep, yu2023detecting, pinna2023investigation, aiken2023measuring, Zhu2023SCGRU:Oversampling, Sabbah2023Self-admittedEmbeddings}\textsuperscript{,}
        
        \citep{skryseth2023technical, gong2023identifying, li2024large, gu2024self, gama2024towards, phaithoon2021fixme, qu2022empirical, karmakar2022experience, sutoyo2024deep}\textsuperscript{,}
        
        \citep{kaur2024embracing, sheikhaei2024empirical, shahzeidi2025hybrid, shivashankar2025beacon}
         & - & 
         No. \citep{
            Dai2017DetectingTrackers,
            Maldonado2017UsingDebt,
            Flisar2018EnhancedIdentification,
            Wattanakriengkrai2019IdentifyingIDF,
            Huang2018IdentifyingMining,
            Liu2018SATDTool}\textsuperscript{,}
            
            \cite {Wattanakriengkrai2019AutomaticIDF,
            Flisar2019IdentificationEmbedding,
            Maipradit2020AutomatedDebt,
            Rantala2020PredictingSelection,
            Rantala2020PrevalenceKL-SATD,
            Maipradit2020WaitDebt}\textsuperscript{,}
            
            \cite {Rajalakshmi2021ClassificationAnalytics,
            Sala2021DebtHunter:Debt,
            Xiao2022CharacterizingSystems,
            Tu2022DebtFree:Learning,
            Chen2022MulticlassXGBoost,
            ALOMAR2022102693,
            Sharma2022Self-admittedCauses}\textsuperscript{,}
            
            \cite {yin2023two,
            satyaAutomated,
            pinna2023investigation,
            gama2024towards,
            phaithoon2021fixme, kaur2024embracing, sheikhaei2024empirical}\textsuperscript{,}
            
            \cite {shahzeidi2025hybrid} 
            & No. \cite {
            Ren2019NeuralExplainability,
            Wang2020DetectingNetworks,
            Santos2020LongDetection,
            Santos2020Self-AdmittedNetwork,
            Sridharan2021DataDetection,
            Zhu2021DetectingCNN-BiLSTM}\textsuperscript{,}
            
            \cite {Santos2021EvaluatingComments,
            Yu2021UsingDebt,
            Zhuang2022AnDetection,
            Khan2022AutomaticPackages,
            Yin2022DeepDebt,
            YU2022111219}\textsuperscript{,}
            
            \cite {Li2022IdentifyingLearning,
            DiSalle2022PILOT:Debt,
            Li2022Self-admittedNetworks,
            li2023automatic,
            li2023debtviz,
            qu2023deep}\textsuperscript{,}
            
            \cite {yu2023detecting,
            Zhu2023SCGRU:Oversampling,
            gong2023identifying, sutoyo2024deep, sheikhaei2024empirical, shahzeidi2025hybrid}
         & No. \citep{
            aiken2023measuring,
            Sabbah2023Self-admittedEmbeddings,
            skryseth2023technical,
            li2024large,
            gu2024self,
            qu2022empirical}\textsuperscript{,}
            
            \cite {karmakar2022experience, sutoyo2024deep, shivashankar2025beacon}\\
        \bottomrule
    \end{tabular}
    \label{tab:tb_methodology}
\end{table}

Manual approaches significantly depend on human intervention, typically requiring manual annotation of code comments or other textual software artifacts to detect SATD. Manual methods rely solely on heuristic-based techniques. There is no application of ML, DL, or Transformer in the manual category, as these techniques are generally used for more automated processes. These methods are frequently time-consuming but offer high precision in identifying subtle or context-specific technical debt items. For example, Potdar and Shihab \citep{Potdar2014AnDebt} conducted research using source code comments from several open-source projects. They manually classified comments as SATD or non-SATD based on specific comment patterns. 
Similarly, Maldonado and Shihab \citep{Maldonado2015DetectingDebt} employed manual labeling to construct an SATD dataset focused on several specific types of debt, including design debt, requirement debt, documentation debt, test debt, and defect debt. This strategy was critical in establishing baseline datasets that could subsequently be utilized to train automated classifiers \citep{Maldonado2017UsingDebt}.

Semi-automated approaches combine manual review with automated tools to improve efficiency while retaining accuracy. 
Semi-automated methods predominantly rely on heuristic-based techniques, including part-of-speech (POS) tagging \citep{Farias2015AComments}, and textual patterns \citep{Bavota2016ADebt}. While some approaches incorporate ML techniques \citep{Yu2020IdentifyingApproach, Rantala2020PrevalenceKL-SATD}, there is no evidence that DL or Transformer-based models are used in these semi-automated techniques.


Automated approaches are characterized by the use of ML, DL, or Transformer and advanced NLP techniques to fully automate SATD detection process. In order to detect SATD from issue tracking systems, Dai and Kruchten \citep{Dai2017DetectingTrackers} created an automated method that uses an ML algorithm, specifically, a Na\"ive Bayes (NB) classifier, along with other NLP approaches such as TF-IDF and TextRank. This method demonstrated the potential for high scalability and integration into software development processes, with promising results in terms of precision, recall, and F1 score. Recently, Li et al. \citep{li2023automatic} introduced a novel approach to automatically detect SATD from four different sources: source code comments, commit messages, pull requests, and issue tracking systems. The most recent trend in automated SATD detection includes the use of Transformer-based models (e.g., BERT \citep{aiken2023measuring, Sabbah2023Self-admittedEmbeddings, sutoyo2024deep}, RoBERTa \citep{karmakar2022experience, skryseth2023technical}, DistilRoBERTa~\citep{shivashankar2025beacon}, and ChatGPT \citep{li2024large}).


\begin{tcolorbox}[colback=gray!5!white, colframe=black, title=Answer to RQ2, boxrule=0.5pt]
Over the past decade, there has been a visible shift from manual towards fully automated approaches to improve efficiency and scalability in SATD detection.\end{tcolorbox}

\subsection{Results - RQ3: Synthesis of reported effectiveness metrics for SATD detection techniques}
\label{rq-metrics}
As emphasized in Section~\ref{sec:Introduction}, a key contribution of this review is the systematic compilation and categorization of reported performance scores across SATD detection techniques. Rather than conducting new experiments or proposing a benchmarking study, we provide a structured cross-technique perspective on how detection approaches have been reported to perform in the literature.

To address RQ3, we synthesized the performance metrics as reported in the selected studies. It is important to note that the synthesized results are limited to within-project evaluations, typically performed using k-fold cross-validation or random splits on the same dataset or project. While some primary studies also reported cross-project or out-of-distribution evaluations (e.g., training on one project and testing on unseen projects), these were deliberately excluded to ensure comparability across studies. Consequently, the reported precision, recall, and F1 scores should be interpreted as reflecting performance under controlled within-project settings, which often yield higher values than cross-project evaluations and may overestimate real-world generalizability.

Following the approach of prior prominent SLRs~\citep{Hall2012AEngineering,malhotra2015systematic,hosseini2017systematic}, we synthesized evidence from diverse data sources and models by compiling results reported in the primary studies, rather than conducting new experiments or direct comparisons. Because these studies vary in dataset characteristics, size, and benchmarking protocols, the aggregated results should be interpreted as indicative trends rather than definitive rankings. Where possible, we report the range and central tendency of precision, recall, and F1 scores to illustrate how different approaches have performed in their original contexts.

Out of 74 relevant studies, 17 did not report their performance results, and 3 only provided AUC scores. Therefore, 54 out of the 74 articles included in our review reported performance results using at least one of the common evaluation metrics: precision, recall, or F1 score. We compiled all performance results reported in these 54 articles and organized them in a spreadsheet. Rather than focusing solely on the best-performing model from each study, we included the results from all model classifiers. This approach provides a more comprehensive and unbiased summary of SATD detection models. By considering all models, researchers can capture a broader range of performance variability, which is essential for understanding the factors that influence model effectiveness \citep{Hall2012AEngineering}. Limiting the analysis to only the best-performing models could lead to a skewed conclusion, as it overlooks the context in which the models perform and fails to account for suboptimal models that offer important insights into potential limitations or challenges. Moreover, analyzing all models enables a more robust cross-study comparison, allowing for the identification of consistent trends and patterns in model performance across different contexts, datasets, and techniques. This holistic approach strengthens the reliability of the conclusion and ensures its applicability across a broader range of scenarios, rather than being limited to cases of optimal performance.

We used boxplots to visualize the synthesized results because they are useful for showing variation across populations without assuming specific data distributions~\citep{Hall2012AEngineering}. We extracted the reported effectiveness of each technique from the primary studies, focusing on precision, recall, and F1 scores computed from the confusion matrix. The boxplots present the distributions of these metrics across studies and enable comparisons by model-related factors.

\begin{itemize}
\item Precision = $\frac{TP}{TP+FP}$ 
\item Recall = $\frac{TP}{TP+FN}$ 
\item F1 score = $\frac{2*Precision*Recall}{Precision+Recall} = \frac{2*TP}{2*TP+FP+FN}$

\end{itemize}

We acknowledge the potential for misinterpretation when aggregating the results of techniques across various studies without considering the diversity of datasets used. We chose this approach to provide a more comprehensive understanding of each model's performance and versatility across a wide range of contexts and scenarios. This inclusive approach enables the identification of universally effective strategies as well as those that excel under specific conditions, thereby providing a nuanced perspective on the potential applicability, hereby offering a holistic view of their effectiveness in detecting SATD. This methodology aligns with precedents in the field, as reflected in systematic literature reviews such as those by Hall et al. \citep{Hall2012AEngineering} and Malhotra \citep{malhotra2015systematic}. This precedent underscores the validity and relevance of cross-dataset comparisons in extracting meaningful insights that transcend the particularities of individual studies. By synthesizing findings across diverse datasets, this review aims to advance SATD detection techniques and provide valuable guidance for future research and practical applications in software engineering.

To avoid any misleading interpretations, we divide the process of SATD detection into two distinct tasks: \textbf{SATD identification} and \textbf{SATD categorization}. SATD identification aims to determine whether a particular entity or instance is present in the text of a software development artifact and to classify it as SATD or non-SATD. SATD categorization, on the other hand, takes this a step further by not only identifying the presence of SATD but also classifying it into specific types of technical debt. These categories may include labels such as design debt, architecture debt, and requirement debt, based on the linguistic elements identified in the text. 

In the following subsection, we present performance results for each SATD detection task, illustrated through boxplots.


\subsubsection{Performance results of SATD identification}
\begin{figure}[htb!] 
    \centering
    \includegraphics[width=350px]{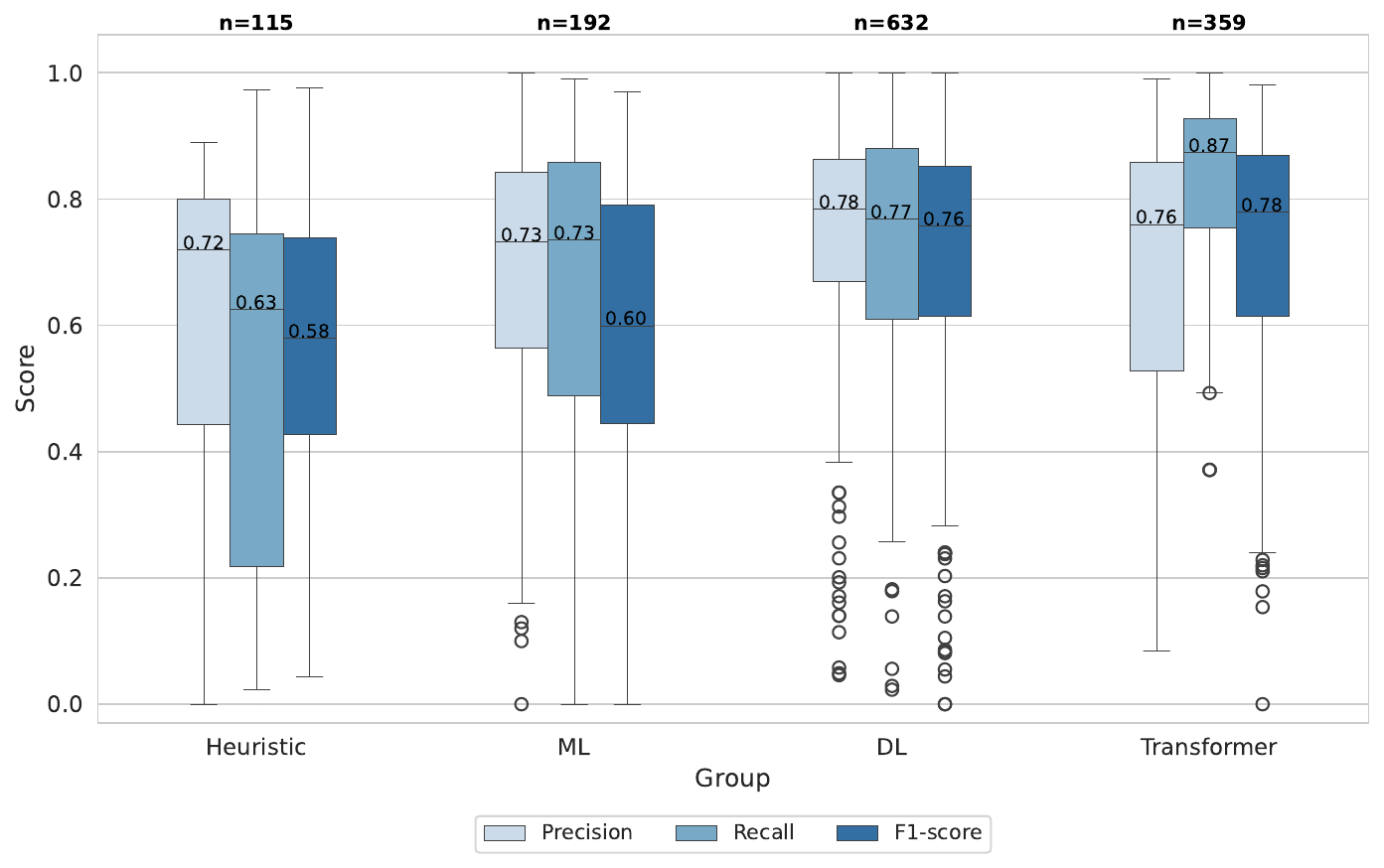}
    \caption{Boxplot of SATD identification grouped by their techniques.}
    \label{fig:group-satd-non-satd}
\end{figure}

The boxplot shown in Figure~\ref{fig:group-satd-non-satd} provides a comparative overview of performance across different SATD identification methods, specifically for binary classification, which aims to accurately distinguish between SATD and non-SATD instances. The three key metrics displayed in the boxplot, namely precision, recall, and F1 score, are standard measures for evaluating the effectiveness of classification models. Precision indicates the model's ability to correctly identify instances as SATD, whereas recall measures how effectively the model captures all relevant SATD instances. The F1 score, which represents the harmonic mean of precision and recall, offers a balanced evaluation, especially when one metric is prioritized over the other. Together, these metrics provide valuable insights into the performance of SATD detection methodologies in identifying technical debt within software projects. Finally, we also report the number of experiments (n) for each technique, giving greater weight to algorithms evaluated more frequently, since larger n values provide more reliable evidence and more stable performance estimates across trials.

The heuristic-based techniques achieve a relatively high precision of 0.72, indicating their ability to correctly identify SATD when classifying an instance as SATD. However, their recall is lower at 0.63, suggesting that they overlook a considerable portion of SATD instances. The F1 score of 0.58 reflects this imbalance, indicating that although heuristic-based methods were foundational in early SATD detection, they now struggle to capture all relevant SATD instances.

In contrast, ML models exhibit more balanced performance, achieving a precision of 0.73 and a recall of 0.73, yielding an F1 score of 0.60. This improvement over heuristic-based methods illustrates ML's effectiveness in reducing false positives while maintaining good coverage of SATD instances. Historically, ML techniques have represented a significant advancement in SATD detection, effectively bridging the gap between precision and recall.

DL models further improve performance, achieving a precision of 0.78 and a recall of 0.77. This yields a balanced F1 score of 0.76. These results suggest that DL techniques are more effective at identifying SATD instances while minimizing false positives. Improvements in both precision and recall demonstrate how DL architectures have advanced SATD identification by enabling more accurate classification across diverse software contexts.

Transformer-based techniques achieved the best overall performance with a precision of 0.76, the highest recall of 0.87, and an F1 score of 0.78. These results highlight the superior ability of Transformer models to identify most SATD instances while maintaining competitive precision. As the latest advancement in SATD detection, Transformers strike a strong balance between minimizing false positives and maximizing the detection of relevant SATD cases.

\begin{figure}[ht!] 
    \centering
    \begin{subfigure}{0.9\textwidth} 
        \centering
        \includegraphics[width=\linewidth]{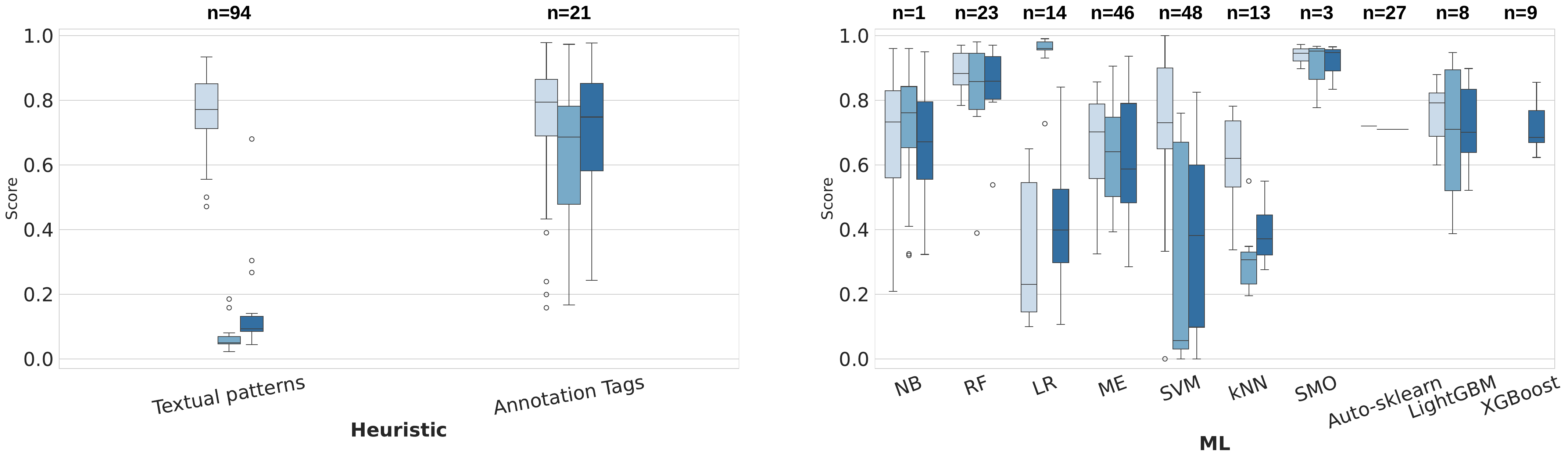} 
        \label{fig:subfigure_binary1}
    \end{subfigure}

    \begin{subfigure}{0.9\textwidth}
        \centering
        \includegraphics[width=\linewidth]{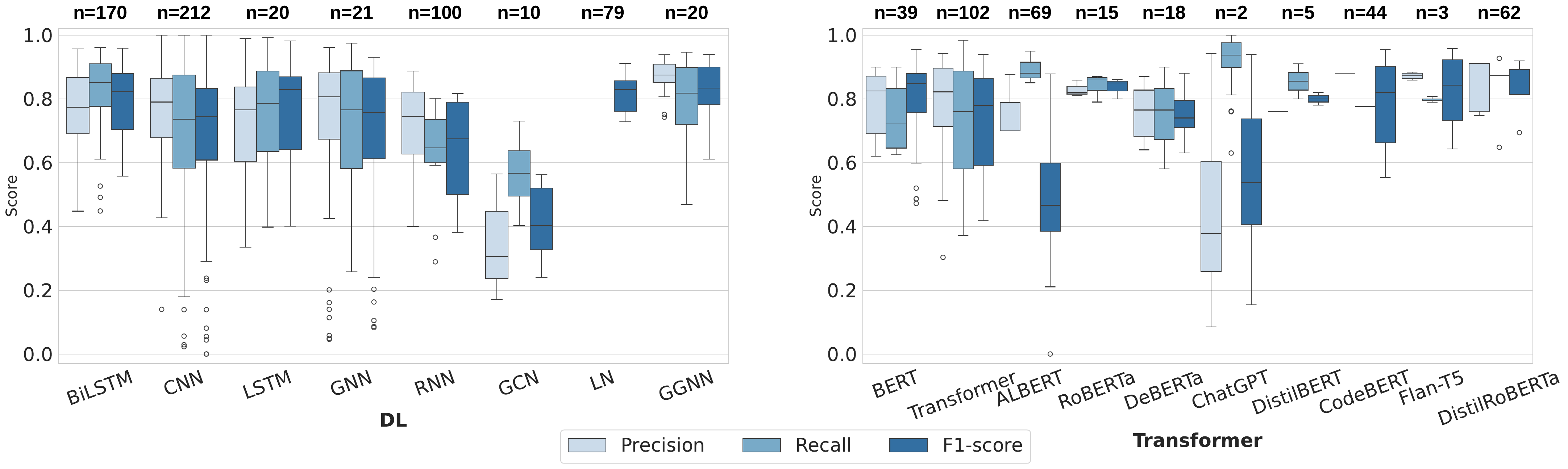} 
        \label{fig:subfigure_binary2}
    \end{subfigure}
    
    \caption{Model performance of SATD identification grouped by each technique.}
    \label{fig:binary-detail}
\end{figure}

Referring to Figure~\ref{fig:group-satd-non-satd}, we can further elaborate and illustrate the metric results in more detail, organizing them based on their respective models, as depicted in Figure~\ref{fig:binary-detail}. Each model is evaluated using a specified number of experiments (denoted by \enquote{n=}), thereby highlighting the range of experiments used to measure its performance.

Heuristic-based techniques, particularly Textual patterns and Annotation tags, were assessed for their effectiveness. Textual patterns (n=94) achieved relatively high precision, indicating that this model accurately classifies SATD examples with low false positives. However, its lower recall and F1 score suggest that it often fails to identify many SATD instances, limiting its overall effectiveness in comprehensive identification. In contrast, Annotation Tags (n=21) performed better in recall and F1 score, indicating a more balanced detection capability, although variability was present, as shown by the wider interquartile range.

ML models exhibited a broader range of performance. Among them, Random Forest (RF) (n=23) provided the most consistent results, achieving high precision and F1 scores with relatively narrow variability. Logistic Regression (LR) (n=14), while reaching high median values, showed substantial variability across experiments, suggesting less stable performance. Maximum Entropy (ME) (n=46) demonstrated moderately consistent outcomes, though with slightly lower scores compared to RF. In contrast, models such as Support Vector Machine (SVM) (n=48) and k-NN (n=13) exhibited lower precision and recall, indicating difficulty in accurately identifying SATD instances. Na\"ive Bayes (NB) (n=1) and Sequential Minimal Optimization (SMO) (n=3) were reported only in a few cases, limiting the generalizability of their results. Notably, Auto-sklearn (n=27), LightGBM (n=8), and XGBoost (n=9) also achieved strong results, although their relatively small sample sizes make it difficult to draw robust conclusions.

DL models show heterogeneous performance in SATD identification. BiLSTM (n=170) consistently delivers strong medians, with recall and F1 generally higher than precision and with moderate variability. CNN (n=212) achieves comparable median scores but exhibits the largest dispersion and many low outliers, indicating less stable results across studies. LSTM (n=20) attains solid medians with noticeable spread, while GNN (n=21) also reaches competitive median values but with wide variability and several low-score cases. In contrast, RNN (n=100) trends lower overall, and GCN (n=10) is the weakest, particularly in precision and F1. Among the stronger DL variants, LN (n=79) and GGNN (n=20) stand out with high median performance; GGNN is especially competitive while remaining reasonably consistent despite its smaller sample size.

Transformer-based models represent the most recent advancements in SATD identification. BERT (n=39) and Transformer (n=102) achieved the strongest and most consistent results, with high precision, recall, and F1 scores. ALBERT (n=69) achieves generally strong precision and recall, but its F1 scores vary widely, with several low outliers, making its results less consistent than BERT and Transformer. RoBERTa (n=15) performed well, though with fewer experiments, while DeBERTa (n=18) showed more variability and somewhat lower scores. ChatGPT (n=2) was evaluated in only a very limited number of cases, yielding inconsistent outcomes. DistilBERT (n=5) and Flan-T5 (n=3) achieved promising results but cannot be generalized due to their small sample sizes. CodeBERT (n=44) and DistilRoBERTa (n=62) demonstrated consistently strong performance, with DistilRoBERTa in particular matching the stability of BERT-based models.


\subsubsection{Performance result SATD categorization}
\begin{figure}[htpb!] 
    \centering
    \includegraphics[width=350px]{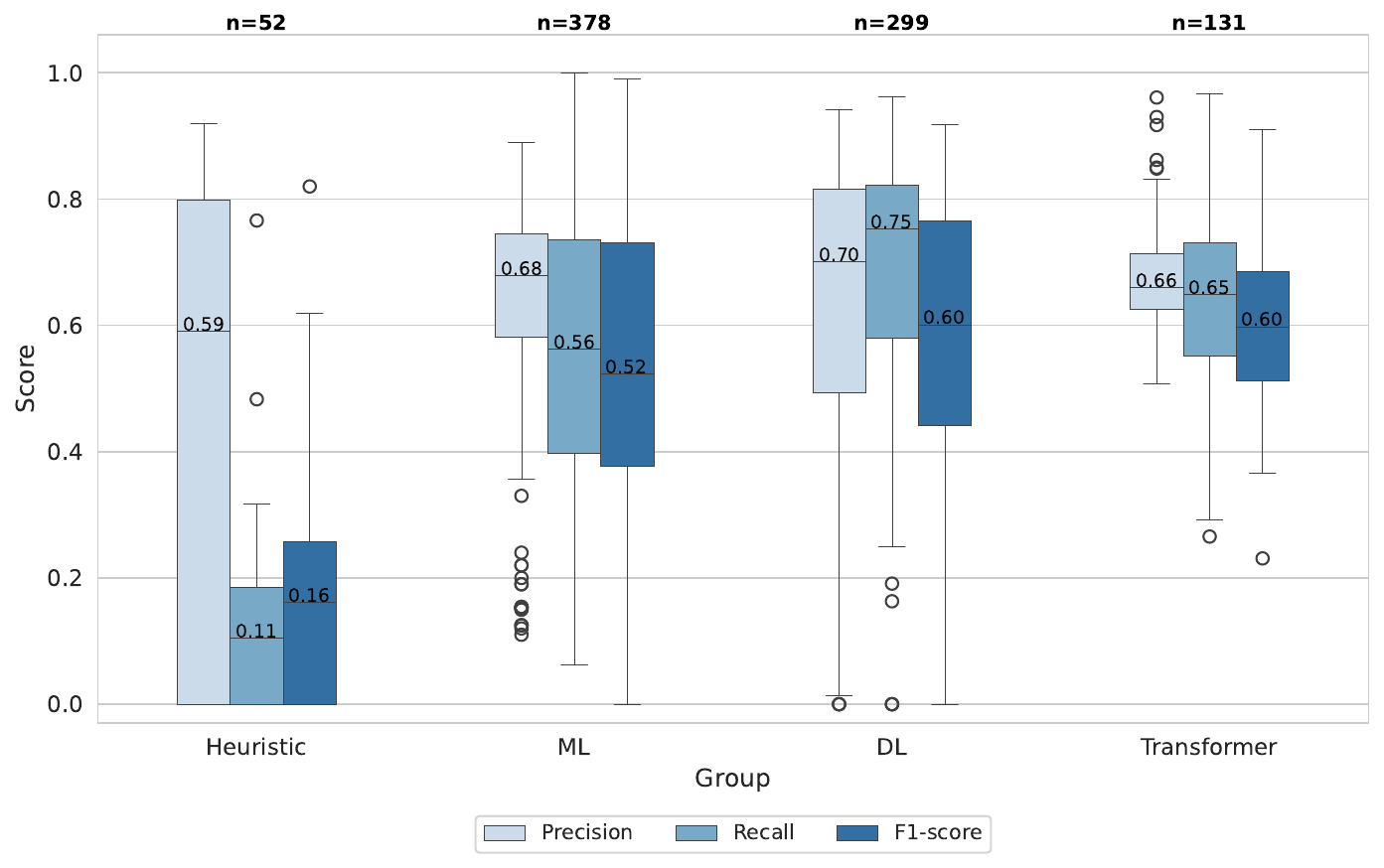}
    \caption{Boxplot of precision, recall, and F1 score of SATD categorization.}
    \label{fig:group-multiclass-satd}
\end{figure}

Figure~\ref{fig:group-multiclass-satd} offers insight into the precision, recall, and F1 score of these techniques, focused on various SATD categorization methodologies for multiclass classification, and provides insight into how effectively these compiled techniques differentiate between types like design debt, requirement debt, documentation debt, test debt, defect debt, and others. The metrics provide insights into how effectively these techniques perform in a multiclass classification scenario, in which distinguishing among different categories of SATD is more complex than in binary classification.

Heuristic-based techniques demonstrate a relatively high precision of 0.59 but a very low recall of 0.11 and an F1 score of 0.16, indicating significant difficulty in capturing true SATD instances and missing many relevant examples. This limitation affects their effectiveness in real-world applications, particularly in complex multiclass classification tasks. In contrast, ML techniques show notable improvement, achieving a precision of 0.68 and a recall of 0.56, resulting in a more balanced F1 score of 0.52. These results suggest that ML models achieve higher accuracy in SATD classification than heuristics, although they still miss a considerable number of true positives, as indicated by the recall.

DL techniques perform well, achieving a precision of 0.70 and a recall of 0.75, yielding an F1 score of 0.59. This indicates that DL models capture most true positives (high recall) while maintaining reasonable precision; however, the overall balance (F1) remains moderate-higher than ML (0.52) but slightly below Transformer (0.61). Transformer-based techniques, despite being more advanced, achieved a precision of 0.66, a recall of 0.65, and an F1 score of 0.61. While their performance is not markedly higher than that of DL in this setting, Transformers demonstrate consistent, balanced results, outperforming heuristic approaches and ML models and providing comparable effectiveness to DL, with a slightly higher F1 score. While Transformer typically excels in complex text classification tasks, these results may reflect the influence of model configuration or data characteristics on its performance in this specific SATD identification task. Nonetheless, Transformer clearly outperforms Heuristic-based techniques, achieves higher overall scores than ML models, and performs on par with DL approaches, with a slightly better F1 score.

\begin{figure}[h!]
    \centering
    \begin{subfigure}{0.9\textwidth} 
        \centering
        \includegraphics[width=\linewidth]{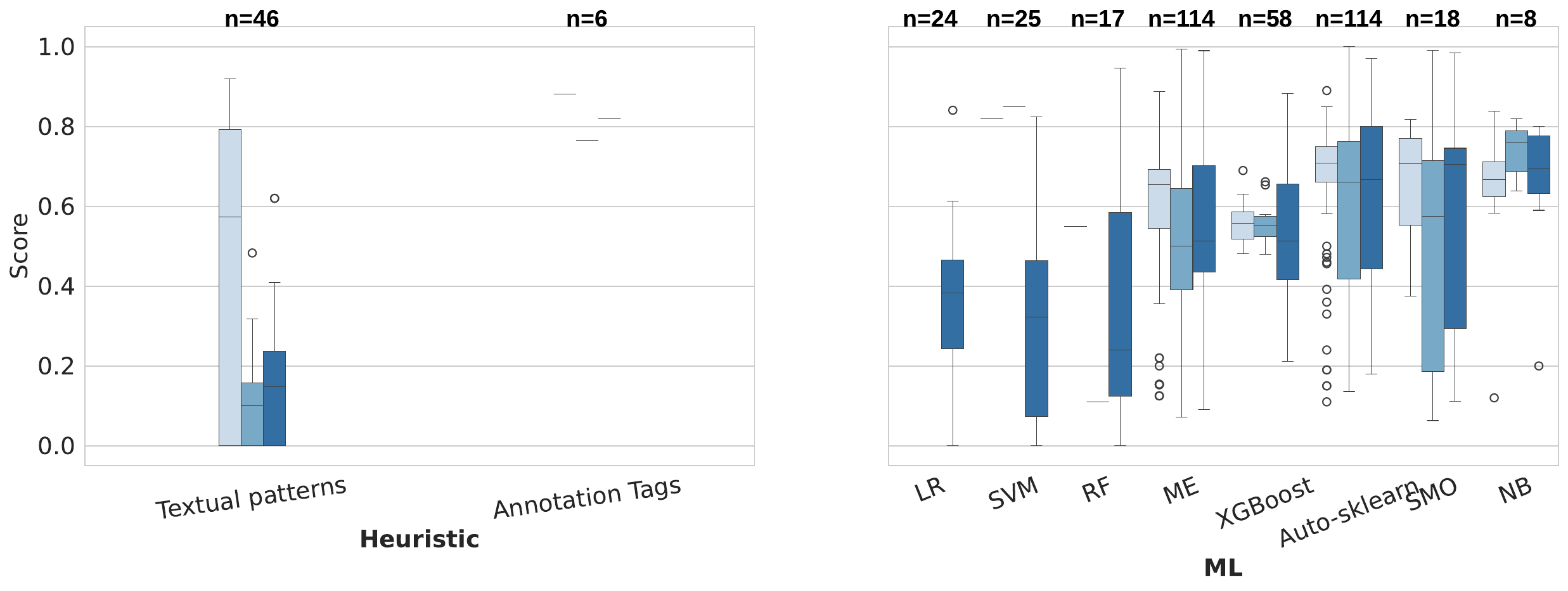} 
        \label{fig:subfigure_multiclass1}
    \end{subfigure}

    \begin{subfigure}{0.9\textwidth}
        \centering
        \includegraphics[width=\linewidth]{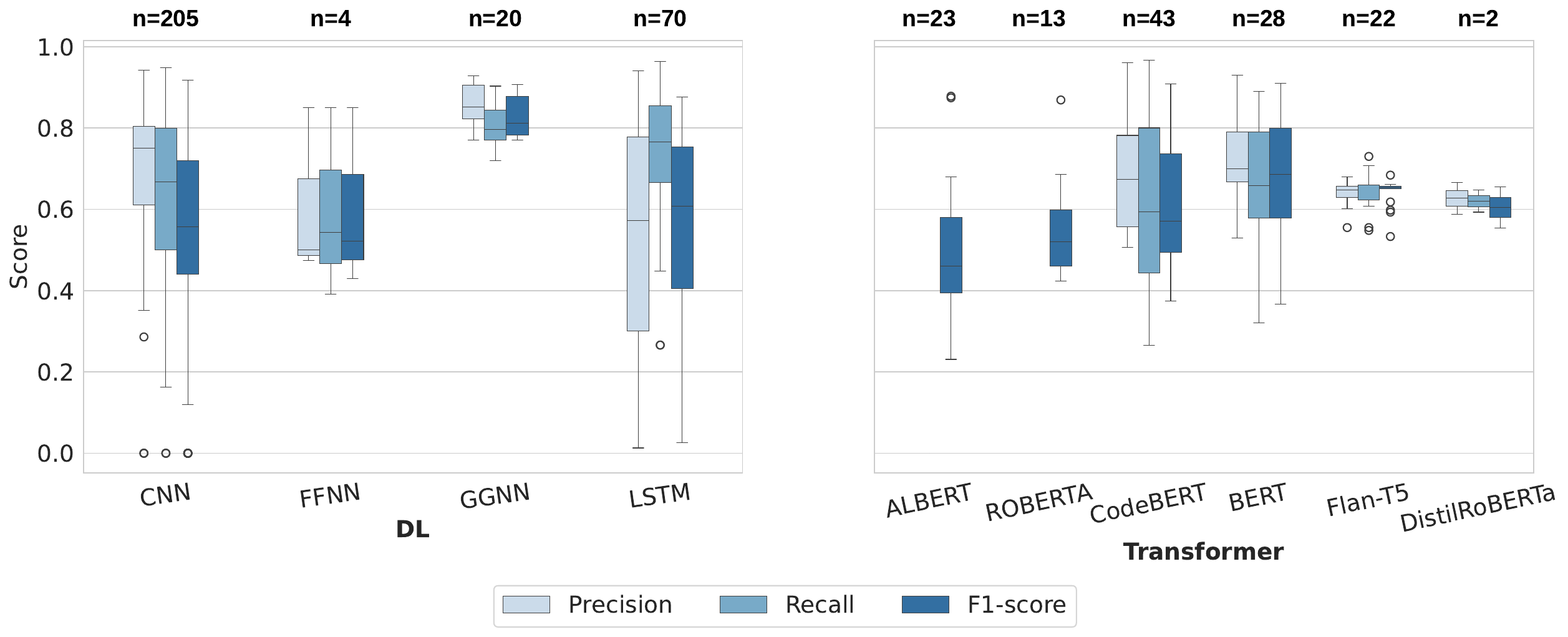} 
        \label{fig:subfigure_multiclass2}
    \end{subfigure}
    
    \caption{Model performance of SATD categorization grouped by each technique.}
    \label{fig:multiclass-detail}
\end{figure}

Referring to Figure~\ref{fig:group-multiclass-satd}, we can provide a comprehensive evaluation of specific types of SATD categorization, organized by their respective models, as shown in Figure~\ref{fig:multiclass-detail}. Each model is evaluated using a specified number of experiments (denoted by \enquote{n=}), thereby highlighting the range of experiments used to measure its performance.

Heuristic-based techniques were evaluated using Textual patterns (n=46) and Annotation tags (n=6). Textual patterns showed highly variable performance, with precision around 0.6 but much lower recall and F1 scores, indicating that while this method can sometimes classify SATDs accurately, it struggles to consistently retrieve all instances. The wide interquartile range and presence of outliers highlight its inconsistency across contexts. In contrast, Annotation tags, although tested in only six cases, showed relatively stable performance, with higher precision, recall, and F1 scores, suggesting stronger performance but limited generalizability. These findings confirm the limitations of heuristic approaches: while sometimes effective for achieving high precision, they often miss critical SATD instances, leading to suboptimal recall and lower overall F1 scores.

ML models exhibited a broad spectrum of results, reflecting their algorithmic diversity. Random Forest (RF) (n=114) and Auto-sklearn (n=114) achieved the strongest performance, with consistently high precision and recall, effectively balancing false positives and missed SATD cases. In contrast, Na\"ive Bayes (n=17) performed more weakly, especially in recall. Other models, such as Support Vector Machine (SVM) (n=25), Logistic Regression (LR) (n=24), and XGBoost (n=58), showed lower recall and greater variability, indicating challenges in reliably categorizing SATD. Maximum Entropy (ME) (n=18) demonstrated moderate performance, whereas Sequential Minimal Optimization (SMO) (n=8) was tested in only a limited number of cases, thereby limiting its generalizability.

DL models demonstrated clear improvements over heuristic and ML techniques. CNN (n=205) achieved strong and consistent results across metrics, while LSTM (n=70) also performed well but with greater variability. Gated Graph Neural Network (GGNN) (n=20) showed promising outcomes, though based on fewer samples. Feedforward Neural Network (FFNN) (n=4) displayed relatively strong performance, but the extremely small sample size limits conclusions about its broader applicability.

Transformer-based models emerged as the most advanced group. BERT (n=28) achieved consistently high precision, recall, and F1 scores, standing out as the top performer. ALBERT (n=23) also performed strongly, though with more variability, while RoBERTa (n=13) maintained high scores despite its smaller sample size. Additional models, including CodeBERT (n=43) and Flan-T5 (n=22), also demonstrated solid and balanced performance, reinforcing their reliability for SATD detection. DistilRoBERTa (n=2), however, was evaluated on very limited data, making its results less conclusive.

From Figures~\ref{fig:binary-detail} and~\ref{fig:multiclass-detail}, we can determine the best-performing algorithm for each group based on precision, recall, and F1 score. It is important to note that Figure~\ref{fig:binary-detail} represents SATD identification aimed at distinguishing SATD from non-SATD (binary classification), while Figure~\ref{fig:multiclass-detail} illustrates the results of SATD categorization of specific SATD types (multiclass classification). To select the best-performing algorithm from the boxplot, we primarily focused on the F1 score (dark blue bars) as it offers a balance between precision and recall. We identified the algorithm with the highest median F1 score, indicated by the middle line of the boxplot, and considered its variance (the spread of the box and whiskers), with lower variance indicating more consistent performance. Additionally, we examined precision and recall (light and mid-blue bars) to ensure a balanced trade-off between the two, prioritizing algorithms with high values for both. In addition, we minimized the influence of algorithms with numerous outliers, as such algorithms indicate inconsistent performance, even when their median scores are high. 


\begin{table}[htpb!]
    \centering
    \caption{Summary of highest performance results based on SATD detection tasks.}
    \begin{tabular}{l>{}m{3.5cm}>{}m{3.5cm}>{}m{4.3cm}}
        \toprule
        \multirow{2}{*}{\textbf{Technique}} & \multicolumn{3}{c}{\textbf{Metric Evaluations}}\\ 
        \cline{2-4}
        & \textbf{Precision} & \textbf{Recall} & \textbf{F1 score}\\
        \hline
        \multicolumn{4}{c}{\textbf{SATD identification}}\\
        \hline
        Heuristic & Annotation tags & Annotation tags & Annotation tags\\
        \hline
        ML & RF & RF & RF \\
        \hline
        DL & CNN & BiLSTM & BiLSTM\\
        \hline
        Transformer & DistilRoBERTa & DistilRoBERTa & DistilRoBERTa \\
        \hline
        \multicolumn{4}{c}{\textbf{SATD categorization}}\\
        \hline
        Heuristic & Annotation tags & Annotation tags & Annotation tags\\
        \hline
        ML & Auto-sklearn & Auto-sklearn & Auto-sklearn\\
        \hline
        DL & GGNN & GGNN & GGNN\\
        \hline
        Transformer & BERT & BERT & BERT\\
        \bottomrule
    \end{tabular}
    \label{tab:tb_best_satd_types}
\end{table}

The best-performing algorithms for each technique and SATD task are summarized in Table~\ref{tab:tb_best_satd_types}. Overall, Transformer-based models consistently deliver the strongest reported results across both SATD identification and categorization. For SATD identification, DistilRoBERTa yields the highest precision, recall, and F1 score among Transformer variants in our synthesis, indicating strong and balanced effectiveness in binary classification settings. ML models, particularly Random Forest, also show competitive performance for identification, especially in precision and F1, while among DL models BiLSTM achieves the strongest overall balance between precision and recall. For SATD categorization, BERT remains the best-performing Transformer model, achieving the highest precision, recall, and F1 score, while Auto-sklearn and GGNN lead the ML and DL groups, respectively. These outcomes reflect trends reported across the literature rather than a definitive ranking, as performance is strongly influenced by dataset quality, annotation practices, and evaluation protocols.

\begin{tcolorbox}[colback=gray!5!white, colframe=black, title=Answer to RQ3, boxrule=0.5pt]
Overall, Transformer-based models achieve the strongest reported effectiveness across SATD detection tasks. In SATD identification, DistilRoBERTa attains the highest precision, recall, and F1 score in the synthesized results, while in SATD categorization BERT is the best-performing model across all three metrics. Although several ML (e.g., RF and Auto-sklearn) and DL models (e.g., BiLSTM and GGNN) also achieve strong results in specific settings, their performance is generally less consistent than Transformer-based techniques across the heterogeneous datasets and evaluation protocols used in the primary studies.
\end{tcolorbox}

\subsection{Results - RQ4: Proposed tools for SATD detection} 
\label{rq-tools}
To address this research question, we conducted a comprehensive review of existing tools developed over the past decade. While SATD detection and management have been extensively studied, there remains a significant gap in translating research findings into practical, developer-friendly tools. This section focuses on tools specifically designed for SATD detection, providing insights into advancements in automation and the various techniques used to address different forms of technical debt.

In this SLR, we classify an approach as a tool only when the authors explicitly propose a prototype or a ready-to-use system, rather than merely providing source code. To be considered a valid contribution to SATD detection, we regard a tool as a system designed for external use, meaning it can be executed with minimal setup and without requiring significant modifications and must be deployable and practically usable \citep{kernighan1976software, se_terminology}. While some studies provide source code to support their experiments, the availability of source code alone is not sufficient to qualify an approach as a tool. Additionally, we include studies in which a tool is described and a replication package is provided, even if the authors do not explicitly refer to it as a tool or if the provided link is no longer valid, as long as the paper provides sufficient evidence that the tool was originally intended for SATD detection.

Table~\ref{tab:tb_tools} presents a collection of SATD detection tools proposed in different software projects. This compilation reflects the evolution of tool development over time, highlighting an increasing emphasis on automation, ML integration, and varied strategies for managing different types of technical debt. The following discussion examines the significance of these tools within the broader context of SATD detection and identifies key trends that have shaped their development over the past decade.

\begin{table}[ht!]
    \centering
    \caption{SATD detection tools proposed.}
    \begin{tabular}{m{2cm}>{}m{0.5cm}>{}m{1.2cm}>{}m{1.4cm}>{}m{1.25cm}>{}m{1.91cm}>{}m{4.4cm}>{}m{1.67cm}}
        \toprule
        \textbf{Name} & \textbf{Year} & \textbf{Technique} & \textbf{Automation Level} & \textbf{Artifact} & \textbf{Task Type} & \textbf{Description} & \textbf{Status}\\

       \midrule
        BEACon-TD \citep{shivashankar2025beacon} & 2025 & Transformer & Automated
 & Issue &  Identification, Categorization & DistilRoBERTa-based tool that classifies software issues as technical debt and further categorizes them into 13 types using a two-stage ensemble of binary classifiers & Available\\



        \midrule
        DebtViz \citep{li2023debtviz} & 2023 & DL & Automated
 & Code comment \& Issue &  Categorization & A tool for SATD detection, classification, visualization, and monitoring in a single platform to categorize code/design, documentation, test, and requirement debt types & Available\\

        \midrule

        SATDBailiff \citep{ALOMAR2022102693} & 2022 & ML & Automated
 & Code comment &  Identification & A tool designed to mine, identify, and track SATD & Available\\
        
        \midrule
        A browser extension \citep{Khan2022AutomaticPackages} & 2022 & ML & Automated
 & Issue &  Categorization & A browser extension utilizes a developed ML model to automatically categorize the SATD types (i.e., documentation, code, defect, test, design, build, architecture, versioning, usability, and requirement debt) of an rOpenSci R package & Not available\\

        \midrule
        DebtHunter \citep{Sala2021DebtHunter:Debt} & 2021 & ML & Automated
 & Code comment &  Identification, Categorization & A machine learning-based technique for detecting self-admitted technical debt & Available\\
        \midrule
        FixMe \citep{phaithoon2021fixme} & 2021 & ML & Automated
 & Code comment &  Categorization & A GitHub bot that is developed to detect, monitor, and notify developers about On-hold SATD in their repositories & Broken link\\

        \midrule   
        SATD Detector \citep{Liu2018SATDTool} & 2018 & ML & Automated
 & Code comment &  Identification & A Java library and an Eclipse plug-in. The tool automatically identifies whether comments are SATD or non-SATD and integrates with an IDE to help manage these comments & Obsolete\\

        \midrule 
        eXcomment \citep{Farias2015AComments, farias2021comment} & 2015 & Heuristic & Semi-Automated
 & Code comment & Identification & A tool designed to parse Java source code and fetch code comments to identify SATD or non-SATD & Broken link\\
        \bottomrule
    \end{tabular}
    \label{tab:tb_tools}
\end{table}

The proposed tools for SATD identification employ diverse techniques and varying levels of automation to address technical debt in software projects. eXcomment \citep{Farias2015AComments, farias2021comment} utilizes a heuristic approach with semi-automated functionality to parse Java source code comments and detect SATD. In contrast, SATD Detector \citep{Liu2018SATDTool} applies ML in a fully automated manner to identify and classify SATD derived from code comments within an IDE, although it has since become obsolete.

Advancing the role of ML, tools such as DebtHunter \citep{Sala2021DebtHunter:Debt} and FixMe \citep{phaithoon2021fixme} implement automated SATD detection and categorization. DebtHunter employs ML to detect SATD, while FixMe functions as a GitHub bot that actively monitors and notifies developers about unresolved On-hold SATD.

SATDBailiff \citep{ALOMAR2022102693} further extends SATD management by introducing an automated tool designed to mine, track, and monitor SATD occurrences over time. Another approach is demonstrated by a browser extension \citep{Khan2022AutomaticPackages}, which leverages an ML model to automatically categorize different SATD types, such as documentation, usability, and requirement debt. Lastly, DebtViz \citep{li2023debtviz} adopts DL to provide an automated solution that integrates detection, classification, visualization, and monitoring of various SATD types within a single platform. More recently, BEACon-TD \citep{shivashankar2025beacon} introduces a Transformer-based solution that detects SATD and classifies it into 13 specific types using a two-stage ensemble of binary classifiers.

This trend of limited tool development is further highlighted when considering that, despite the SLR encompassing 74 relevant studies, only 8 articles proposed specific tools for SATD identification. This disparity highlights a concerning gap between the substantial research interest in SATD and the development of practical, automated solutions to assist developers in managing technical debt. While a range of tools has been introduced, their limited availability indicates that much of the research has yet to translate into actionable, industry-ready solutions. Addressing this gap will be critical for advancing the practical management of technical debt in software projects, moving from theoretical frameworks toward robust, scalable automation.







\begin{tcolorbox}[colback=gray!5!white, colframe=black, title=Answer to RQ4, boxrule=0.5pt]
We identify several tools proposed for SATD detection, with a focus on automation, ML integration, and addressing different types of technical debt. However, despite significant research interest, only a few tools have been developed, revealing a gap between theoretical advancements and the availability of practical, automated solutions for managing technical debt in real-world software projects.
\end{tcolorbox}

\subsection{Results - RQ5: Primary challenges associated with SATD detection}
\label{rq5}
To address the challenges associated with SATD detection, we conducted a thorough analysis of the relevant literature, focusing on the difficulties commonly highlighted by researchers. These challenges reflect the complexities inherent in identifying SATD across various software projects and methodologies. 

We identified nine key challenges related to SATD identification from the relevant studies, as summarized in Table~\ref{tab:tb_challenges}, and each study emphasizes one of these challenges. The extraction process focused on statements reported in the \enquote*{Discussion} or \enquote*{Threats to Validity} sections of the reviewed papers. All discernible challenges, whether explicitly or implicitly mentioned, were collected for further analysis.

The grouping of challenges was also performed using a bottom-up thematic analysis. Each extracted challenge statement was coded and then clustered into higher-level categories based on thematic similarity, following common practices in prior SE literature reviews. The first author conducted the initial coding and grouping, while the second author independently cross-checked a 15\% random sample of the classifications using the shared spreadsheet. Any disagreements were resolved through collaborative discussion until a consensus was reached. Only one case required clarification, indicating that disagreements were minimal and consistently resolvable. Each resulting challenge category is explained in this section and further elaborated on below.

\begin{table}[htb!]
    \centering
    \caption{Papers highlighting key challenges. Studies that fall into more than one challenge are highlighted with square brackets.}
    \begin{tabular}{l>{}m{4.96cm}>{}m{11.1cm}>{}m{8cm}}
        \toprule
        \textbf{No} & \textbf{Challenge} & \textbf{Paper Reference}\\
        \midrule
        1 & Error-proneness  & No. \citep{Potdar2014AnDebt, Bavota2016ADebt}\textsuperscript{,[}\citep{Maldonado2015DetectingDebt,Farias2015AComments,Dai2017DetectingTrackers,Wattanakriengkrai2019IdentifyingIDF,Xavier2020BeyondSystems,Li2020IdentificationTrackers,qu2022we}\textsuperscript{]}
        \\

        \midrule
        
        2 & Imbalanced data & No. \textsuperscript{[}\citep{Huang2018IdentifyingMining,Ren2019NeuralExplainability,Wang2020DetectingNetworks,Farias2020IdentifyingVocabulary,Santos2020LongDetection,Rantala2020PredictingSelection,Rantala2020PrevalenceKL-SATD,Santos2020Self-AdmittedNetwork,Sala2021DebtHunter:Debt,Santos2021EvaluatingComments,Yu2021UsingDebt, Zhuang2022AnDetection,Azuma2022AnDockerfile,Yin2022DeepDebt,YU2022111219,DiSalle2022PILOT:Debt,yin2023two,aiken2023measuring,Zhu2023SCGRU:Oversampling,gong2023identifying,gama2024towards}\textsuperscript{,} \citep{
        Wattanakriengkrai2019AutomaticIDF,
        Sridharan2021DataDetection, 
        Zhu2021DetectingCNN-BiLSTM, 
        Chen2022MulticlassXGBoost, 
        Li2022Self-admittedNetworks, 
        Russo2022WeakSATD:Debt, 
        sheikhaei2023automated, 
        yu2023detecting}\textsuperscript{],}
        
        \citep{Sabbah2023Self-admittedEmbeddings,skryseth2023technical,qu2022empirical,kaur2024embracing,ishimoto2024empirical}
        \\
        \midrule   

        3 & Generalizability & No. \textsuperscript{[}\citep{Dai2017DetectingTrackers,Wattanakriengkrai2019IdentifyingIDF,Xavier2020BeyondSystems,Li2020IdentificationTrackers,Maldonado2015DetectingDebt,Liu2018SATDTool, Ren2019NeuralExplainability,Farias2020IdentifyingVocabulary,Rantala2020PredictingSelection,Rantala2020PrevalenceKL-SATD,Sala2021DebtHunter:Debt,Wang2020DetectingNetworks,Santos2021EvaluatingComments,Azuma2022AnDockerfile,Yin2022DeepDebt,YU2022111219,DiSalle2022PILOT:Debt,yin2023two,aiken2023measuring,Zhu2023SCGRU:Oversampling,gong2023identifying,qu2022we, li2024large, gama2024towards}\textsuperscript{],}
        
        \citep{Maipradit2020AutomatedDebt, Khan2022AutomaticPackages, Tu2022DebtFree:Learning, Li2022IdentifyingLearning, Yu2020IdentifyingApproach, satyaAutomated, sutoyo2024deep, shahzeidi2025hybrid, shivashankar2025beacon}
         \\

        \midrule   

        4 & Explainability & No. 
        \textsuperscript{[}\citep{Ren2019NeuralExplainability,Wang2020DetectingNetworks,Zhuang2022AnDetection,Yin2022DeepDebt,YU2022111219,DiSalle2022PILOT:Debt,li2024large,Guo2021HowStudy,gu2024self}\textsuperscript{]}\\

        \midrule  
        
        5 & False negatives & No. 
        \textsuperscript{[}\citep{Huang2018IdentifyingMining,Santos2020LongDetection,Santos2020Self-AdmittedNetwork}\textsuperscript{]}\\


        \midrule  
        
        6 & Risk of reproducibility & No. 
        \textsuperscript{[} \citep{Guo2021HowStudy,gu2024self,ALOMAR2022102693}\textsuperscript{]}
        \\

        \midrule  
        
        7 & Computational costs & No. 
        \textsuperscript{[}\citep{Guo2021HowStudy,Yu2021UsingDebt,Zhuang2022AnDetection,ALOMAR2022102693,gu2024self}\textsuperscript{],} \citep{sheikhaei2024empirical}\\

        \midrule  
        
        8 & Overlap of features among debt types & No. 
        \textsuperscript{[}\citep{Maldonado2015DetectingDebt, Chen2022MulticlassXGBoost}\textsuperscript{]}\\

        \midrule  
        
        9 & Tagging and comment relevance & No. 
        \textsuperscript{[}\citep{Farias2015AComments,qu2022we}\textsuperscript{]}\\
        
        \bottomrule
    \end{tabular}
    \label{tab:tb_challenges}
\end{table}

\begin{description}
    \item[Error proneness] 

The first challenge is the dependence on manual analysis, which introduces the potential for human error and inconsistency \citep{Bavota2016ADebt}. Since the process relies on human judgment, it is prone to inconsistencies in labeling, as different individuals may interpret and categorize SATD in different ways~\citep{Farias2015AComments}. This subjectivity in analysis can lead to varying outcomes in tasks such as pattern recognition and feature engineering, thereby reducing the reliability of the process.

The variability in human interpretation introduces a lack of uniformity in the labeled datasets, which can significantly affect the effectiveness of ML models that rely on this data for training. When models are trained on datasets that contain inaccuracies or inconsistencies, their predictive accuracy and generalization capabilities suffer. This challenge becomes even more pronounced when scaling up to larger datasets, where consistent and precise labeling is critical to maintaining the quality of the training data. Without robust and reliable analysis, models may struggle to handle the increased complexity of large datasets, resulting in reduced performance and misclassification of SATD instances.

\item[Imbalanced data]
In most software projects, SATD comments represent only a small portion of the total code comments, resulting in a significant class imbalance. This imbalance can adversely affect the performance of SATD identification models, as they may focus on non-SATD comments that dominate the dataset. As a result, the models may struggle to accurately identify the relatively rare SATD instances, often prioritizing the majority class and overlooking the minority class \citep{Chen2022MulticlassXGBoost}.

This imbalance frequently leads to a high number of false negatives \citep{he2009learning}, in which the model fails to detect SATD instances, thereby undermining its effectiveness. Since ML models generally find it difficult to learn patterns from underrepresented classes \citep{krawczyk2016learning}, the scarcity of SATD examples makes it challenging for these models to develop a robust understanding of SATD characteristics. Consequently, the model's ability to identify and classify SATD instances diminishes, particularly when faced with a large volume of non-SATD items. Addressing this imbalance is critical to improving the precision and reliability of SATD identification models, as it ensures that models are not biased toward detecting the majority class while neglecting crucial SATD instances.

\item[Generalizability]
One of the primary challenges in identifying SATD is its generalizability. Generalizability refers to the extent to which SATD identification models maintain their accuracy and effectiveness when applied across different software projects and domains. This challenge arises from the diverse nature of software projects, which often exhibit significant variation in coding practices, domain-specific terminology, and architectural designs \citep{Guo2021HowStudy}.

Models trained on data from one specific project tend to be tailored to the characteristics of that project, limiting their performance when applied to other projects with different practices. For example, the way SATD is expressed in comments may vary across projects due to differences in the development team's language, domain-specific jargon, or coding style conventions \citep{Maipradit2020AutomatedDebt, Li2020IdentificationTrackers}. These project-specific factors can result in models that are highly effective within their training environment but struggle to generalize to other contexts, especially when encountering unfamiliar coding practices or new domains.


\item[Explainability]
Another primary challenge in the SATD identification process, particularly in ML and DL-based models, is the lack of explainability. This issue arises because many advanced models, such as neural networks and DL architectures, function as \enquote{black boxes,} meaning their internal decision-making processes are not easily interpretable by users \citep{emmert2020explainable}. These models generate predictions without providing transparent explanations about the underlying reasoning for classifying an instance as SATD, non-SATD, or even specific types of SATD.

The absence of explainability presents a significant obstacle to the adoption and trustworthiness of SATD detection tools. Moreover, explainability is crucial for validating the accuracy of SATD identification models. If developers cannot trace the logic behind a model's predictions, it becomes difficult to diagnose errors or refine the model to reduce false positives or false negatives.

\item[False negatives]
False negatives are another concern in SATD detection. They occur when SATD items are misclassified as non-SATD, resulting in undetected technical debt in the codebase. This occurs when models fail to identify instances in which developers have either implicitly or explicitly admitted technical debt \citep{Bavota2016ADebt}. As a result, undetected technical debt can linger in the system, leading to long-term code degradation, higher maintenance costs, and an increased risk of introducing more bugs and system failures \citep{tom2013exploration, lim2012balancing}.

\item[Risk of reproducibility]
Another primary challenge in the context of machine learning and supervised model-based models is the issue of reproducibility \citep{Guo2021HowStudy, ALOMAR2022102693}. These models often involve complex architectures with numerous parameters that must be fine-tuned for optimal performance. However, many studies fail to provide sufficient details on the hyperparameters, training configurations, or the specific processes used to achieve their results. Additionally, the lack of publicly available code exacerbates reproducibility issues, as other researchers cannot replicate the exact experimental setup. This makes it difficult to reproduce results across different studies and datasets, raising concerns about the reliability and validity of these techniques.

\item[Computational costs]
Another challenge associated with SATD detection is the high computational costs of many advanced models, particularly DL and Transformer techniques. These models require significant resources during training, including high-performance GPUs, large memory capacities, and extended computation time. This high computational cost poses several obstacles for researchers and practitioners who aim to implement or improve SATD identification methods.

Additionally, the extended computation time required for training DL models presents a challenge in terms of practicality. SATD identification models often require retraining or fine-tuning to adapt to new projects or datasets, and the time-consuming nature of this process can hinder agile software development practices. In scenarios where rapid feedback is essential, such as continuous integration and deployment environments, the lengthy training times of DL models can become a bottleneck, delaying the detection of technical debt and the subsequent mitigation efforts.

\item[Overlap of features among debt types]
Furthermore, the overlap of features among different types of technical debt, such as code debt, design debt, and requirement debt, presents another challenge \citep{Chen2022MulticlassXGBoost}. This overlap makes distinguishing between different debt types difficult, leading to potential misclassification. The ambiguity in distinguishing between SATD and other debt types can reduce the accuracy of identification models. If features that are common to multiple debt types are not carefully handled, models may struggle to identify SATD correctly.

\item[Tagging and comment relevance]
Finally, tagging and comment relevance are a critical challenge in identifying SATD. Developers often use tags such as \enquote{TODO,} \enquote{FIXME,} or \enquote{XXX} in code comments or other textual artifacts to mark areas that require attention or indicate temporary fixes. While these task-related tags are commonly employed as indicators of SATD, they are not always relevant to technical debt. In many cases, tags such as \enquote{TODO} or \enquote{FIXME} can serve as simple reminders or non-technical debt items, thereby creating noise in the identification process. Moreover, not all SATD instances are explicitly marked with such tags, further complicating the identification process.
\end{description}

The results highlight significant trends in SATD detection, demonstrating an evolution from heuristic-based methods to advanced machine learning and deep learning techniques. However, these findings raise important questions regarding the applicability, generalizability, and practical adoption of these approaches. The following section presents a critical discussion of our results, compares them with the existing literature, and outlines their implications for future research and industry practice.

\begin{tcolorbox}[colback=gray!5!white, colframe=black, title=Answer to RQ5, boxrule=0.5pt]
We identified nine major challenges in detecting SATD, including manual analysis errors, data imbalance, and limited generalizability and explainability in advanced models. Other challenges include false negatives, reproducibility risks, computational costs, overlapping features among debt types, and the relevance of tagging and comments, all of which hinder effective SATD identification.
\end{tcolorbox}

\section{Discussion}
\label{section_discussion}
The findings from this systematic review reveal several key trends and challenges in detecting SATD over the past decade, demonstrating both the evolution and fragmentation within the field. SATD detection techniques have progressed significantly, shifting from simple heuristic-based methods to more complex ML, DL, and Transformer-based models. While each advancement has brought increased performance and automation, the development of practical and scalable SATD detection tools remains limited.

\subsection{Evolution of SATD detection techniques and artifacts}
\label{subsection-evolution}
Early SATD detection techniques primarily relied on heuristic-based methods, such as keyword matching and textual patterns, to identify technical debt in code comments. These techniques, although foundational, were often limited by their inability to account for the nuanced language used by developers and their reliance on rule-based systems that could not easily adapt to different contexts. From 2017 onward, ML techniques such as NB, SVM, and RF became more prevalent, offering improved accuracy by learning from labeled datasets and reducing false positives. Despite this progress, these methods still faced challenges related to generalizability across different datasets and projects, as well as difficulty in explaining the rationale behind the classification of certain comments as SATD.

The rise of DL methods from 2020 to 2022 marked a significant leap forward in SATD detection. Models like LSTM and CNN enabled the identification of more complex patterns in developer comments and project artifacts. These techniques improved recall and reduced false negatives, thereby providing a more balanced technique for identifying SATD across diverse datasets. However, the high computational cost of training and deploying these models remains a major barrier to widespread adoption.

Most recently, Transformer-based models, such as BERT and RoBERTa, have emerged as the leading techniques for SATD detection. These models have demonstrated superior performance in capturing the context and semantics of developer comments, enabling more accurate and generalizable SATD identification. The ability of Transformer to handle large-scale datasets and complex text has pushed the boundaries of what is possible in SATD detection. However, issues related to the explainability of these models and their scalability for industrial use still persist.

It is also important to note that the evolution of these techniques has occurred primarily within the context of code comments, which remain the dominant artifact for SATD detection. While this focus is understandable given the accessibility and ubiquity of comments across software projects, it narrows the diversity and generalizability of existing approaches. Other artifacts, such as issue trackers, commit messages, and pull requests, have been comparatively underexplored despite their potential to provide complementary perspectives on technical debt. Addressing this imbalance represents a valuable direction for future research, as expanding the scope of artifacts can enable a more comprehensive understanding of SATD across the software development process.

\subsection{Challenges in SATD detection}
\label{subsection-challenges}
Despite the advancements in SATD detection techniques, several challenges hinder the practical adoption of these methods in real-world software development environments. A key challenge in SATD detection is the imbalance in datasets—SATD instances are often much less frequent than non-SATD instances, which negatively impacts model performance. This imbalance is not limited to the distinction between SATD and non-SATD but also occurs among specific SATD types, where certain categories, such as requirements debt or design debt, being far less represented, further complicating detection. Although several techniques have been employed to address data imbalance in SATD detection, future research should explore more advanced methods, such as leveraging large language models (LLMs) to better handle imbalanced data and improve model performance. 

Generalizability remains another key challenge. Many SATD detection models are trained on specific datasets, often from open-source projects, which limits their performance on new or proprietary codebases. This lack of generalizability reduces the practical utility of these models in diverse software development environments, where coding practices, project documentation, and domain-specific language can vary significantly. 

To address this, future research should move beyond within-project validation and systematically evaluate SATD detection models \enquote{in the wild}, using out-of-distribution datasets and cross-project settings. Such evaluations, as demonstrated by recent work like \citep{shivashankar2025beacon}, are essential for understanding how well these models transfer to unseen contexts and for ensuring their reliability in real-world adoption.

The explainability of advanced models, particularly DL and Transformer-based techniques, is also a concern. Without a clear understanding of why certain texts are classified as SATD, developers may be reluctant to trust and adopt these models in their workflows. Future work should focus on developing explainable AI techniques for SATD detection to ensure that these models' decision-making processes are transparent and understandable to developers. Research on enhancing the interpretability of DL and Transformer models, perhaps through visualizations or simplified intermediate representations, would help practitioners better trust and utilize these models in real-world software development settings.

Another significant barrier to adoption is the high computational cost associated with training and maintaining DL and Transformer-based models. While these models offer improved accuracy, their resource-intensive nature makes them difficult to implement for smaller development teams or in environments with limited computational resources. This challenge highlights the need for more efficient algorithms or alternative techniques to deliver high accuracy without the associated computational burden.

\subsection{Lack of tools for SATD detection}
\label{subsection-lack-tools}
Despite the significant academic interest in SATD detection, the translation of research into practical tools remains limited. Over the past decade, this systematic review identified only seven tools developed specifically for SATD detection. This reflects a notable gap between theoretical research and practical application and a lack of sufficient tools to meet the diverse needs of software development teams. While tools such as SATD Detector and eXcomment provide valuable functionality for identifying SATD in Java codebases, their scope is limited in terms of flexibility, scalability, and integration with modern software development workflows. Consequently, these tools have experienced limited adoption beyond academic prototypes and small-scale studies. The current state of SATD detection tools presents several challenges that hinder their widespread adoption among practitioners. Key areas of concern include limited real-time integration, inadequate support for multiple programming languages, insufficient collaboration features, and poor scalability for large codebases.

Modern software development practices are heavily reliant on continuous integration (CI) and continuous deployment (CD) pipelines \citep{zhao2017impact}, alongside integrated development environments (IDEs) \citep{alizadehsani2022modern} such as IntelliJ IDEA, Visual Studio Code, and Eclipse. However, most SATD detection tools lack seamless integration with these environments, necessitating manual execution or batch processing. This disrupts established workflows and reduces the tool's utility for practitioners who require real-time feedback during development. The absence of real-time notifications limits the capacity of developers to address technical debt in a timely manner, thereby contributing to its accumulation. Tools designed for practical use should therefore provide IDE integration and support for automated scanning within CI/CD pipelines to enhance workflow efficiency and developer productivity.

Furthermore, existing tools often struggle to scale effectively in large and complex software projects. Many of these tools are designed with a narrow focus on specific programming languages, particularly Java, which limits their applicability in diverse development environments where teams work with multiple languages, such as Python, C++, and JavaScript. This restriction reduces their practicality for organizations seeking comprehensive SATD detection across their entire codebase. Performance optimization for large repositories and multi-language compatibility are thus critical areas for future improvement. In addition, many SATD detection tools rely on ML or DL models, which are often perceived as \enquote{black boxes}. This lack of interpretability may make it challenging for developers to fully trust and adopt these tools. Offering more transparent and explainable outputs—such as highlighted patterns, relevant keywords, or confidence scores—could enhance clarity, build trust, and encourage broader adoption among practitioners.


Moreover, existing tools seldom offer functionality for long-term monitoring and trend analysis of SATD. For projects with extended lifecycles, it is valuable to track how technical debt is introduced, accumulated, and eventually repaid. While tools such as DebtViz aim to visualize and monitor SATD over time, they remain in early stages of development and may not yet be suitable for large-scale applications. Enhancing these tools with features for historical analysis and prioritization of technical debt repayment could support more informed decision-making and strategic refactoring efforts.

These limitations highlight a broader challenge: the gap between academic research and industry needs. Many tools are developed with a strong focus on academic performance metrics, yet they may not fully address practical constraints, including integration, scalability, and usability. Strengthening collaboration between academia and industry is crucial to ensuring that SATD detection tools are not only scientifically rigorous but also practical and widely applicable in real-world software development settings.


\subsection{Decision-making guidance for adopting SATD detection techniques}
\label{subsection-decision-making}
To address the lack of actionable guidance for adoption, this subsection synthesizes evidence from our review into practical decision considerations for selecting SATD detection techniques. Below, we summarize the findings based on the techniques, approaches, or models that were identified in our study. We also summarize the main advantages and disadvantages of each technique or approach.

\begin{itemize}
  \item \textbf{Heuristic-based techniques} (e.g., tagging rules, regular expressions, pattern matching) are most appropriate when teams need a lightweight solution with minimal infrastructure and strong interpretability. They are typically easy to implement, explain, and integrate into existing workflows, making them suitable for small-to-medium codebases or for quick triage (e.g., flagging obvious TODO/FIXME-like debt markers). However, the review evidence indicates that heuristics often miss more subtle SATD expressions and are sensitive to vocabulary differences across projects and developer writing styles. Therefore, heuristics are best positioned as a baseline or as a high-precision filter when false positives are costly, rather than as a comprehensive detector.\\
  - \textbf{Advantages}: Lightweight, easy to use
  \\
  - \textbf{Disadvantages}: Lower recall, limited coverage

  \item \textbf{Classic ML approaches} (e.g., SVM, RF, and LR) offer a pragmatic middle ground for teams that have access to labeled data and seek improvements over heuristics without the full operational burden of deep models. Across the reviewed studies, ML methods tend to perform more robustly than heuristics while remaining easier to reproduce and debug than DL/Transformer approaches. ML is often a sensible choice for organizations that can maintain a modest labeling pipeline and require reasonably transparent features (e.g., TF-IDF-based models) to secure stakeholder buy-in. Nevertheless, reported performance remains highly dependent on dataset characteristics (e.g., class imbalance, artifact type, and preprocessing), and practitioners should expect re-tuning when moving to new domains.\\
  - \textbf{Advantages}: More robust than heuristics, can be reasonably interpretable (feature-based)
  \\
  - \textbf{Disadvantages}: Requires labeled data, performance may degrade under domain shift

  \item \textbf{Deep learning approaches} (e.g., CNN/LSTM/GNN) are most appropriate when the goal is improved effectiveness under complex linguistic variation or when large training corpora are available. The review evidence shows that DL methods often improve detection quality relative to ML in within-project evaluations, but they entail higher operational costs (training time, hardware, maintenance) and reduced transparency. DL models are therefore best suited to teams that can sustain the MLOps overhead, have sufficient data volume, and accept that model explanations may require additional techniques (e.g., post hoc explainability) to support developers' trust.\\
  - \textbf{Advantages}: Often higher effectiveness than classic ML, can leverage larger corpora and richer representations
  \\
  - \textbf{Disadvantages}: Higher compute and MLOps cost, lower transparency

  \item \textbf{Transformer-based models} (e.g., BERT-family and related variants) are typically justified when SATD detection is a high-value, high-scale activity and teams prioritize effectiveness and broader language understanding. In the reviewed literature, Transformers often achieve the strongest reported results for SATD identification and are increasingly used across diverse textual artifacts. The main practical constraints are operational: they require more computational resources, more specialized expertise, and careful engineering for deployment and monitoring. In addition, their \enquote{black-box} nature can hinder adoption unless paired with interpretability and documentation practices. Practitioners should treat Transformers as a strategic investment: most suitable for large projects, multi-artifact pipelines, or scenarios in which missing SATD instances (false negatives) incur high downstream costs.\\
  - \textbf{Advantages}: State-of-the-art effectiveness, strong language understanding across artifacts
  \\
  - \textbf{Disadvantages}: Highest compute and expertise requirements, interpretability and deployment complexity
\end{itemize}

These qualitative observations are consistent with the overall performance patterns synthesized in Section~\ref{rq-metrics}, where more recent DL and Transformer-based methods tend to report higher precision/recall/F1 than earlier heuristic and classical ML approaches, albeit under heterogeneous datasets and evaluation protocols.

Our synthesis indicates that no single technique dominates across all criteria; the best choice depends on project constraints (e.g., available labeled data, compute budget, and expertise) and the intended use of the detector (e.g., quick triage vs.\ high recall detection). Based on the cross-study synthesis, practitioners can use the following cues to narrow down their choice: if the goal is immediate, low-cost triage with maximum explainability, \textbf{heuristic-based techniques} are typically the best starting point. If labeled data are available and reproducibility and maintainability are important, \textbf{classic ML} provides a strong and practical baseline. If performance requirements are higher and sufficient data/compute resources exist, \textbf{DL techniques} become appropriate. Finally, if an organization can support the operational overhead and needs state-of-the-art effectiveness across heterogeneous artifacts, \textbf{Transformer-based models} are generally the most suitable option.

\subsection{Implications for practitioners}
\label{subsection-implications}
BERT-based models demonstrate superior precision and recall in SATD detection: by adopting these models, practitioners can significantly reduce their reliance on manual review processes, which are often time-consuming, inconsistent, and prone to human error. Automated approaches streamline the identification of technical debt, ensuring that critical SATD instances are accurately detected without the extensive overhead of manual analysis. This enables software teams to allocate their resources more efficiently and to focus on higher-value activities, such as refactoring and feature development.

However, one of the key challenges in SATD identification is class imbalance, where SATD instances are significantly outnumbered by non-SATD instances. This imbalance often leads detection models to favor the dominant (non-SATD) class, resulting in missed detections. Additionally, class imbalance affects SATD categorization, as SATD must be classified into specific types of technical debt, such as design debt, architecture debt, requirements debt, and others. In both cases, the imbalance increases the risk of false negatives \citep{he2009learning}, where important SATD instances are overlooked. To mitigate this issue, practitioners should employ strategies such as data augmentation (e.g., by creating additional synthetic SATD samples to balance the dataset) and domain-adaptation techniques, which optimize model performance across diverse projects and environments. These approaches help ensure that models maintain high accuracy and recall, even when faced with the imbalanced data typical of real-world software projects.


As outlined in subsection~\ref{subsection-decision-making}, we present a comprehensive framework to assist practitioners in selecting the most appropriate SATD detection approach tailored to their specific project needs and constraints. This framework enables practitioners to effectively balance trade-offs involving scalability, explainability, and resource demands. These factors are essential for the successful adoption of SATD detection tools.

Additionally, automated SATD detection and early identification of technical debt contribute to the sustainability of software systems. Since SATD is prevalent across many software projects (affecting between 2.4\% and 31\% of files \citep{Potdar2014AnDebt}), it presents a long-term challenge. SATD often persists for extended periods, with a median lifespan ranging from 18 to 172 days and, in some cases, lasting through more than 1,000 commits \citep{DaMaldonado2017AnDebt}. By adopting automated detection techniques, organizations that adopt these techniques will likely see measurable improvements in development efficiency, reduced technical debt accumulation, and enhanced product quality. These benefits help mitigate the risks associated with delayed debt repayment.\\





    

\subsection{Future directions}
\label{subsection-future-direction}
To address the challenges identified in this review, future research should prioritize enhancing the generalizability, explainability, and efficiency of SATD detection models. Techniques such as transfer learning, which allow models to be trained on one dataset and then applied to others with minimal retraining, could significantly improve generalizability. At the same time, incorporating explainable AI (XAI) techniques is crucial to ensure that the decision-making processes of these models are transparent and understandable for developers. Developing more interpretable models or employing post-hoc explanation techniques could foster trust in these tools and encourage their adoption in real-world projects. Future work should focus on improving the interpretability of deep learning and Transformer models, for example, by using visualizations, highlighting informative patterns, and providing simplified intermediate representations. These improvements would help practitioners better understand the models, build trust in their outputs, and use them more effectively in software development contexts.

Beyond improving models, future research should prioritize SATD detection tools that are scalable, easy to use, and designed for real-world adoption. In practice, this means tools that integrate smoothly with existing development environments, support team collaboration, and work across diverse software ecosystems, including multi-language codebases. Since modern workflows are centered on IDEs and CI/CD pipelines, SATD detectors should be deployable in both settings and provide timely, actionable signals. In particular, real-time or near-real-time feedback can help developers identify and address technical debt as part of their normal workflow, rather than as a follow-up activity performed after implementation. By closing the gap between research prototypes and production-ready tooling, the software engineering community can ensure that SATD detection techniques deliver tangible value for long-term software maintenance and evolution.

While our findings provide valuable insights into SATD detection, it is essential to acknowledge potential limitations that may influence the generalizability of our conclusions. The next section discusses threats to validity, including biases in study selection, dataset heterogeneity, and variations in evaluation metrics, and how we have addressed these concerns.

\section{Threats to Validity}
\label{section_threats}
As with any systematic literature review, the results of this study are subject to various validity threats. These are categorized as internal, external, and construct validity threats, along with mitigation strategies to minimize their impact.

\subsection{Internal validity}

Internal validity refers to the reliability of the methodology employed in the study. One key threat to internal validity in this SLR is related to the formulation of the search strings. In the pilot search, several iterations of search strings were tested, starting with a more general query (e.g., \enquote{self-admitted technical debt} OR SATD and NLP), which yielded a broader, less focused set of results. Over time, more refined strings were developed, including terms such as \enquote{detection} and \enquote{identification}, as well as the use of Boolean operators and the wildcard (*) for broader term variations.

The final search string (S3), which combined terms such as \enquote{self-admitted technical debt} OR SATD with NLP OR \enquote{natural language processing} AND (detection OR identification OR prediction), was chosen after multiple pilot tests to ensure it captured the most relevant studies. This iterative process helped mitigate the risk of missing key studies during the initial literature search.

Furthermore, data extraction inaccuracies could threaten the reliability of our findings. To reduce this risk, the data extraction process was double-checked, and disagreements were resolved through discussions between the authors. A pilot data extraction was also conducted to further minimize potential bias.

\subsection{External validity}
External validity refers to the extent to which the findings of this SLR can be generalized to other contexts beyond the studies reviewed. A primary threat to external validity in this research is the limited generalizability of the results due to the scope of the reviewed datasets. All studies included in this review rely on open-source projects, which may not fully capture the complexities or constraints encountered in proprietary or large-scale industrial environments.

Open-source datasets often have characteristics, such as well-documented codebases or community-driven development practices, that differ significantly from proprietary software projects, which may involve stricter deadlines, unique domain requirements, or higher security and compliance constraints. As a result, some of the trends and findings related to SATD detection may not extend to these more complex and regulated environments. For example, the performance of ML or DL models, which may perform well on open-source datasets, could degrade when applied to more diverse or proprietary datasets with varying development practices.

Another potential limitation of this review is that the synthesized results for RQ3 are based exclusively on within-project evaluations (e.g., k-fold cross-validation or random splits on the same project). While this choice ensures comparability across studies, it may overestimate performance relative to cross-project or out-of-distribution evaluations, which better reflect real-world deployment scenarios. Some primary studies did report cross-project results, but these were not included in our synthesis. Therefore, the performance metrics presented in this review should be interpreted with caution, as they represent results under controlled within-project settings rather than true generalizability across projects.

\subsection{Construct validity}

Construct validity concerns whether a study accurately measures the concepts it aims to investigate. A significant threat in this SLR is the comparison of results across studies, which can be misleading if studies employ different methodologies or report metrics inconsistently. Directly comparing the precision, recall, and F1 scores of models trained on heterogeneous datasets without considering their context may lead to inaccurate conclusions.

To address this, we employed a grouping strategy to compare similar studies and ensured that the SATD tasks were organized before any comparisons were made. By contextualizing the results within groups, we aimed to make the comparisons more meaningful. Furthermore, this methodology is based on established practices within the field, exemplified by Hall et al. \citep{Hall2012AEngineering} and Malhotra \citep{malhotra2015systematic}, which similarly compare results across studies to derive broad-based insights into algorithmic performance across varying contexts.

Having considered the potential limitations of this study, we now summarize our key findings and contributions. The final section revisits the research questions, highlights the main takeaways, and outlines directions for future work to advance SATD detection research and tool development.

\section{Conclusion}
\label{section_conclusion}
This SLR provides an in-depth analysis of SATD detection techniques over the past decade, shedding light on the evolution of techniques from traditional keyword-based methods to sophisticated ML, DL, and Transformer-based models. Among the key findings, it is evident that while heuristic-based techniques were foundational in early SATD detection, they struggle to achieve comparable recall and overall performance to ML and DL techniques. Transformer-based models, particularly BERT, have emerged as the most effective for capturing complex patterns in textual artifacts, achieving superior performance across multiple metrics.

However, several challenges persist, including computational costs, model scalability, and the explainability of AI techniques. The reliance on open-source datasets in many studies limits the generalizability of findings, and further research is needed to adapt these models to diverse industrial environments. Moreover, the imbalance in SATD datasets continues to affect model robustness, and integrating multiple textual artifacts, such as commit messages and pull requests, introduces further complexity. Furthermore, the lack of widely adopted tools for SATD detection in the industry highlights a gap between research advancements and practical applications.

In conclusion, while the field of SATD detection has advanced considerably, further research and collaboration between academia and industry are needed to ensure that these advancements translate into practical tools that enhance software maintainability and reduce the long-term impact of technical debt.

\bmsection*{ORCID}
Edi Sutoyo \orcidlink{0000-0002-8413-5070} \hyperlink{https://orcid.org/0000-0002-8413-5070}{https://orcid.org/0000-0002-8413-5070}\\
Andrea Capiluppi \orcidlink{0000-0001-9469-6050} \hyperlink{https://orcid.org/0000-0001-9469-6050}{https://orcid.org/0000-0001-9469-6050}



\bmsection*{Data availability}
Supplementary material related to this article, including all the boxplots as images, can be found online at \href{https://github.com/edisutoyo/satd-detection-slr}{https://github.com/edisutoyo/satd-detection-slr}

\bmsection*{Declaration of competing interest}
The authors declare that they have no known competing financial interests or personal relationships that could have appeared to influence the work reported in this paper.

\bmsection*{Acknowledgement}
This work was financially supported by the Indonesian Education Scholarship (BPI) from the Center for Higher Education Funding and Assessment (PPAPT), Indonesia Endowment Fund for Education (LPDP), and the Ministry of Higher Education, Science, and Technology of the Republic of Indonesia.





\bibliography{Main}







\end{document}